\documentclass[sigconf]{acmart}
\AtBeginDocument{%
  }

\setcopyright{acmlicensed}
\copyrightyear{2025}
\acmYear{2025}
\setcopyright{acmlicensed}\acmConference[ICMR '25]{Proceedings of the 2025 International Conference on Multimedia Retrieval}{June 30-July 3, 2025}{Chicago, IL, USA}
\acmBooktitle{Proceedings of the 2025 International Conference on Multimedia Retrieval (ICMR '25), June 30-July 3, 2025, Chicago, IL, USA}
\acmDOI{10.1145/3731715.3733411}
\acmISBN{979-8-4007-1877-9/2025/06}

\usepackage{xcolor}
\newcommand{\red}[1]{\textcolor{red}{#1}}
\usepackage{subcaption}
\usepackage{graphicx}
\usepackage{overpic}
\usepackage{dashrule}  
\usepackage{tikz}
\begin{document}

\title{OT-Talk: Animating 3D Talking Head with Optimal Transportation}

\author{Xinmu Wang}
\authornote{This work was done during an internship at Futurewei Techonologies.}
\email{xinmuwang@cs.stonybrook.edu}
\affiliation{%
  \institution{Stony Brook University}
  \city{Stony Brook}
  \state{NY}
  \country{USA}
}

\author{Xiang Gao}
\email{gao2@cs.stonybrook.edu}
\affiliation{%
  \institution{Stony Brook University}
  \city{Stony Brook}
  \state{NY}
  \country{USA}
}

\author{Xiyun Song}
\email{xsong@futurewei.com}
\affiliation{%
  \institution{Futurewei Technologies}
  \city{Santa Clara}
  \state{CA}
  \country{}
}

\author{Heather Yu}
\email{hyu@futurewei.com}
\affiliation{%
  \institution{Futurewei Technologies}
  \city{Santa Clara}
  \state{CA}
  \country{USA}
}

\author{Zongfang Lin}
\email{lin@futurewei.com}
\affiliation{%
  \institution{Futurewei Technologies}
  \city{Santa Clara}
  \state{CA}
  \country{USA}
}

\author{Liang Peng}
\email{lpeng@futurewei.com}
\affiliation{%
  \institution{Futurewei Technologies}
  \city{Santa Clara}
  \state{CA}
  \country{USA}
}

\author{Xianfeng Gu}
\email{gu@cs.stonybrook.edu}
\affiliation{%
  \institution{Stony Brook University}
  \city{Stony Brook}
  \state{NY}
  \country{USA}
}
\renewcommand{\shortauthors}{Xinmu Wang et al.}

\begin{abstract}
Animating 3D head meshes using audio inputs has significant applications in AR/VR, gaming, and entertainment through 3D avatars. However, bridging the modality gap between speech signals and facial dynamics remains a challenge, often resulting in incorrect lip syncing and unnatural facial movements. To address this, we propose OT-Talk, the first approach to leverage optimal transportation to optimize the learning model in talking head animation. Building on existing learning frameworks, we utilize a pre-trained Hubert model to extract audio features and a transformer model to process temporal sequences. Unlike previous methods that focus solely on vertex coordinates or displacements, we introduce Chebyshev Graph Convolution to extract geometric features from triangulated meshes. To measure mesh dissimilarities, we go beyond traditional mesh reconstruction errors and velocity differences between adjacent frames. Instead, we represent meshes as probability measures and approximate their surfaces. This allows us to leverage the sliced Wasserstein distance for modeling mesh variations. This approach facilitates the learning of smooth and accurate facial motions, resulting in coherent and natural facial animations. Our experiments on two public audio-mesh datasets demonstrate that our method outperforms state-of-the-art techniques both quantitatively and qualitatively in terms of mesh reconstruction accuracy and temporal alignment. In addition, we conducted a user perception study with 20 volunteers to further assess the effectiveness of our approach.
\end{abstract}   



\maketitle

\section{Introduction}
\label{sec:intro}

Speech-driven head/face animation is a pivotal task in computer vision and graphics. Due to the massive availability of 2D audio-visual datasets, extensive deep learning-based studies \cite{suwajanakorn2017synthesizing, thies2020neural, alghamdi2022talking} have been conducted to produce 2D talking head videos. But they cannot be directly used for AR/VR or 3D films or games. Reconstructing 3D face/head shapes from 2D monocular videos \cite{habibie2021learning, pham2017speech} is limited by the accuracy of 3D reconstruction techniques. With the release of several audio-mesh datasets of 3D talking head, e.g., VOCASET \cite{cudeiro2019capture}, BiWi \cite{fanelli20103}, and Multiface \cite{wuu2022multiface}, recent studies begin to animate 3D head geometry based on well-captured audio and 3D scanned data. Most of these studies \cite{richard2021meshtalk, cudeiro2019capture, xing2023codetalker, fan2022faceformer} try to learn the mapping from audio signals to facial dynamics, i.e., a encoder to embed the audio features and a decoder to vertex movements. After extracting audio features using techniques like DeepSpeech \cite{hannun2014deep}, wav2vec \cite{baevski2020wav2vec}, Hubert \cite{hsu2021hubert} etc, they estimate the vertex displacement vectors (3D offsets) while keeping their connectivity in the template topology. Although straightforward, directly predicting the 3D offset of each vertex can result in less smooth movements. If the displacement vectors are not predicted smoothly, the facial motion may appear to be jumpy or unnatural, lacking continuity in the deformation process. In addition, displacement-based methods focus on local mesh changes. Although it can describe localized deformations, it can fail to preserve the overall global structure of the face, leading to unrealistic animations. Another drawback of many previous studies is the lack of geometric features, which can be learned from well-structured triangulations. 

To solve the above problems, we propose OT-Talk. We adopt existing encoder-decoder architecture, Hubert-based audio processing module, and transformer-based auto-regressive model. Our contributions lie in 3 aspects. 
\begin{itemize}
    \item  First, to the best of our knowledge, we are the first to apply optimal transportation (OT) in talking head animation. We transform meshes into probability measures and use sliced Wasserstein distance to capture mesh difference to learn smooth and natural facial motions.
    \item Second, we apply a Chebyshev Graph Convolution Network (ChebNet) \cite{defferrard2016convolutional} to learn a geometric-aware head mesh representation, i.e., we try to learn mesh features which are ignored in many previous studies.
    \item  Finally, we conduct an extensive experiment on two public audio-mesh datasets, VOCASET and Multiface, and both quantitative and qualitative results demonstrate the superiority of our approach over baselines.
\end{itemize} In addition, we recruited 20 volunteers to watch the synthesized audio-mesh videos and rate the quality. User perception study also shows that our results are more favored.  

\section{Related Work}
\label{sec:related work}

\subsection{Learning-based Speech-driven 3D Head}

Extensive research \cite{chen2020talking, das2020speech, ji2021audio, prajwal2020lip, vougioukas2020realistic, zhou2021pose} has focused on 2D talking head generation. Due to the lack of available 3D datasets and the difficulty in learning complex 3D geometry, rule-based methods \cite{edwards2016jali, taylor2012dynamic} have been explored for 3D models, utilizing viseme or facial action coding systems. These artist-friendly methods require significant manual effort and frequent adjustments by animators to ensure accuracy, limiting the scalability and efficiency. With the advent of mesh-audio datasets, deep learning methods have made significant strides. \cite{karras2017audio} employs an end-to-end CNN to map audios to 3D vertices, requiring 3-5 minutes of high-quality animation data to replicate a specific actor's talking style. Visemenet \cite{zhou2018visemenet} focuses on generating animator-centric speech motion curves that drive JALI or standard FACS-based face rigs. However, these methods are limited to a single character. EmoTalk \cite{peng2023emotalk} generates accurate blendshape coefficients by separating speech from emotion. However, blendshapes often miss high-frequency facial motion details and are restricted by less interpretable definitions. 

The research most relevant to ours includes VOCA \cite{cudeiro2019capture}, which enables the animation of arbitrary neutral head meshes from speech. FaceFormer \cite{fan2022faceformer} introduces the use of Transformers for this task. CodeTalker \cite{xing2023codetalker} addresses the problem of over-smoothing by using a codebook query operation. SelfTalk \cite{peng2023selftalk} employs a lip-reading interpreter to maintain the consistency of text content. MeshTalk \cite{richard2021meshtalk} attempts to disentangle speech-driven and speech-independent facial movements for the upper and lower face but assumes that such separation holds universally and requires additional expression signals. EmoTalk \cite{peng2023emotalk} improves on MeshTalk by disentangling speech content from emotion through cross-reconstruction, but demands sentences spoken with various emotions, complicating data collection. ScanTalk \cite{nocentini2025scantalk} designs a DiffusionNet-based model that can train on multiple mesh datasets with different mesh structures. FaceDiffuser \cite{stan2023facediffuser} and DiffPoseTalk \cite{sun2024diffposetalk} try to exploit Diffusion model \cite{ho2020denoising} for many-to-many mappings.

\subsection{Geometric Representation for Head Meshes}

Some studies have explored the geometric aspects of 3D head meshes to improve results. E.g., \cite{li2024pose} introduces a geometry-guided audio-vertices attention module to predict vertex displacements with vertex connectivity and hierarchical audio features. \cite{wu2023speech} investigates composite and regional facial movements. However, neither fully utilizes the structure or topology of the 3D mesh. Since head meshes can be represented as graphs, graph convolution networks (GCNs) \cite{bouritsas2019neural, ranjan2018generating, litany2018deformable, cheng2019meshgan} are increasingly used to model and represent 3D head geometry, and \cite{cheng2020faster, li2019generating} achieve high-quality 3D head reconstruction or generation. For dynamic talking heads, GDPnet \cite{liu2021geometry} employs a multi-column GCN (MCGCN-based) mesh encoder to guide the audio encoder in learning latent representations. Inspired by CoMA \cite{ranjan2018generating},  which demonstrated that a head mesh can be represented with a small bunch of parameters in latent space by ChebNet, our work applies a similar method for mesh representation and uses that representation for further deformation.

\subsection{OT for Mesh Deformation}
Representing complex objects via distributions, which are then compared through optimal transport distances, is an active research area to evaluate the dissimilarity between two meshes. Wasserstein distance \cite{peyre2019computational, villani2009optimal} has been widely recognized as an effective optimal transport metric to compare two probability measures, especially when their supports are disjoint. Wasserstein distance has high computational complexity ($O(m^3\log m)$ for time complexity and $O(m^2)$ for memory complexity) and becomes even more problematic in mesh applications where a mesh is treated as a probability measure. Adding entropic regularization and using the Sinkhorn algorithm \cite{cuturi2013sinkhorn} can obtain $\epsilon$-approximation of the Wasserstein distance with a time complexity of $O(m^2/\epsilon^2)$. But the entropic regularization approach doesn't reduce memory complexity and cannot lead to a valid metric between probability measures since the discrepancy does not satisfy triangle inequality. The Sliced Wasserstein distance (SWD) \cite{bonneel2015sliced}, which is computed as the expectation of the Wasserstein distance between random one-dimensional push-forward measures from two original measures, can be solved in $O(m\log m)$ time complexity with a linear memory complexity $O(m)$. Hence, our head animation model is optimized via SWD by encoding the head mesh to a probability measure. 

The plugin estimator \cite{bernton2019parameter} provides a computationally efficient and stable approach to approximate the Wasserstein distance by transforming a mesh into an empirical probability measure. To have a richer representation of the mesh, a discrete probability measure via varifold has been exploited by \cite{glaunes2004diffeomorphic, villani2009optimal, charon2013varifold, kaltenmark2017general}, which is also adopted by our approach for the head animation task.

\section{OT-Talk}

\textbf{Problem Formulation.}
Given two inputs: a head mesh template $\overline{\mathcal{G}}0$ and a sequence of audio snippets $\textbf{A}_{1:T} = (a_1, a_2, ..., a_T)$ with duration $\textbf{T}$, our goal is to predict a sequence of head meshes $\hat{\textbf{G}}_{1:T} = (\hat{\mathcal{G}}_1, \hat{\mathcal{G}}_2, ..., \hat{\mathcal{G}}_T)$ that corresponds to a reference sequence of meshes $\textbf{G}_{1:T} = (\mathcal{G}_1, \mathcal{G}_2, ..., \mathcal{G}_T)$. These predicted meshes have the same number of vertices, faces, and topology as the head template provided by the dataset. We use an encoder-decoder pipeline for this task, with a pretrained Hubert \cite{hsu2021hubert} model for audio encoding. The process for the head meshes is detailed in the following sections.

\subsection{ChebNet Mesh Embedding}

A 3D head mesh is represented as an undirected and connected graph $\mathcal{G} = (\mathcal{V}, \mathcal{E}, \mathcal{W})$, where $\mathcal{V} \in \mathbb{R}^{N \times 3}$ denotes the set of $N$ vertices, $\mathcal{E}$ represents the set of edges, and $\mathcal{W} \in \{0, 1\}^{N \times N}$ is the adjacency matrix indicating the edge connections. The combinatorial graph Laplacian is defined as $\mathcal{L} = \mathcal{D} - \mathcal{W}$, where $\mathcal{D} \in \mathbb{R}^{N \times N}$ is the diagonal matrix with $\mathcal{D}_{ii} = \sum_j \mathcal{W}_{ij}$. The normalized graph Laplacian is given by $\mathcal{L}' = \mathcal{I}_N - \mathcal{D}^{-1/2}\mathcal{W}\mathcal{D}^{-1/2}$, where $\mathcal{I}_N$ is the identity matrix. Since $\mathcal{L}'$ is a real symmetric positive semi-definite matrix, it has a complete set of orthonormal eigenvectors. Therefore, $\mathcal{L}'$ can be diagonalized using the Fourier bases $U = [u_0, u_1, ..., u_{N-1}] \in \mathbb{R}^{N \times N}$, such that $\mathcal{L}' = U \Lambda U^T$, where $\Lambda = \text{diag}[\lambda_0, \lambda_1, ..., \lambda_{N-1}] \in \mathbb{R}^{N \times N}$. The graph Fourier transform for mesh vertices $x \in \mathbb{R}^{N \times 3}$ is defined as $\hat{x} = U^T x$, with the inverse transform given by $x = U \hat{x}$. The convolution operator in the Fourier domain is defined as

\begin{equation}
    x * y = U \left((U^T x) \odot (U^T y)\right)
\end{equation}
which can be computationally expensive when dealing with a large number of vertices, especially since $U$ is not sparse.

The Chebyshev spectral CNN addresses this issue by formulating mesh filtering with a kernel $g_\theta$ parameterized as a recursive Chebyshev polynomial \cite{defferrard2016convolutional} of order $K$:
\begin{equation}
    g_\theta (\Lambda) = \sum_{k=0}^{K-1} \theta_k T_k (\widetilde{\Lambda})
\end{equation}
where $\theta \in \mathbb{R}^K$ represents the Chebyshev coefficient vector, and $T_k$ denotes the Chebyshev polynomial of order $k$ evaluated at the scaled Laplacian $\widetilde{\Lambda} = \frac{2\Lambda}{\lambda_{max}} - \mathcal{I}_N$. The Chebyshev polynomials $T_k$ are computed recursively as $T_k(x) = 2x T_{k-1}(x) - T_{k-2}(x)$ with $T_0 = 1$ and $T_1 = x$. The convolution operation can then be expressed as:
\begin{equation}
    y_j = \sum_{i=1}^{F_{in}} g_{\theta_{i, j}}(L) x_i
\end{equation}
where $y_j$ is the output feature map, $x_i$ is the input feature map, and $\theta_{i, j} \in \mathbb{R}^K$ are the trainable parameters.

To capture both global and local contexts, we resample the mesh during the pooling operation to create hierarchical multi-scale mesh representations. This approach allows convolutional kernels to capture local details in the shallower layers and global context in the deeper layers of the network. We adopt a mesh resampling method similar to that in CoMA \cite{ranjan2018generating}, utilizing transform matrices \( Q_d \in \{0, 1\}^{n \times m} \) and \( Q_u \in \mathbb{R}^{m \times n} \) for in-network downsampling and upsampling, respectively, where \( m > n \). The downsampling matrices are derived by iteratively contracting vertex pairs while minimizing surface error using quadric error metrics (QEM) \cite{garland1997surface}. For a given mesh \( M_i \) and its downsampled version \( M_{i+1} \), we compute the upsampling transform matrix from \( M_{i+1} \) back to \( M_i \) as follows. For a vertex \( \tilde{v}_q \) in mesh \( M_i \), we locate the nearest triangle \( (i, j, k) \) in \( M_{i+1} \) and then calculate the barycentric coordinates, \( \tilde{v}_q = w_i v_i + w_j v_j + w_k v_k \), with \( w_i + w_j + w_k = 1 \). The weights in \( Q_u \) are assigned as \( Q_u(q, i) = w_i \), \( Q_u(q, j) = w_j \), and \( Q_u(q, k) = w_k \), while \( Q_u(q, l) = 0 \) for all other entries.

\subsection{Optimal Transport Metric}

To compare the dissimilarity of predicted mesh and ground truth, we first transform the mesh to a discrete probability measure through varifold. Let triangulation mesh $M$ approximates the smooth shape $X$ which is a sub-manifold of dimension 2 embedded in the ambient space of $\mathbb{R}^n$, where n = 3 for surface, with finite total volume $V(X)<\infty$. For every point $q\in X$, there exists a tangent space $T_qX$ be a linear subspace of $\mathbb{R}^n$. It is essential to orient the tangent space $T_qX$ for every $q$ to establish an orientation of $X$ which ensures that each oriented tangent space can be represented as an element of an oriented Grassmannian. Then $X$ can be associated as an oriented varifold (a distribution on the position and tangent space orientation $\mathbb{R}^n \times \mathbb{S}^{n-1}$) as: $\tilde{\mu} = \int_{X}\delta_{(q, \vec{n}(q))}d V(q)$, where $\tilde{n}(q)$ is the unit oriented normal vector to the surface at $q$. Once established the oriented varifold for a smooth surface, the oriented varifold for $M$ which can be derived by:
\begin{equation}
    \tilde{\mu}^M = \sum_{i=1}^{|F|}\tilde{\mu}^{f_i} = \sum_{i=1}^{|F|}\int_{f_i}\delta_{(q, \vec{n}(q))}d V(q) \approx \sum_{i=1}^{|F|} \alpha_i\delta_{(q, \vec{n}(q))}
\end{equation}
where $q_i$ is the barycenter of the vertices of face $f_i$ and $\alpha_i:=V(f_i)$ is the area of the triangle. We normalize $\alpha_i'$ (sum up to 1) to guarantee that $\tilde{\mu}^M$ possesses the characteristic of a discrete measure.

The Wasserstein-$p$ \cite{villani2021topics} distance for comparing two probability measures $\mu\in \mathbf{P}_p(\mathbb{R}^d)$ and $\nu\in \mathbf{P}_p(\mathbb{R}^d)$ is defined as follows:
\begin{equation}
    W_p^p(\mu, \nu):=\inf_{\pi\in \Pi(\mu, \nu)}\int_{\mathbb{R}_d\times\mathbb{R}_d}||x-y||^p_pd\pi(x, y)
\end{equation}
where $\Pi(\mu, \nu)$ is the set of joint distributions that have marginals are $\mu$ and $\nu$ respectively. It can handle 2 measures with disjointed supports. 

The sliced Wasserstein distance \cite{bonneel2015sliced} (SWD) between two probability measures $\mu \in \mathbf{P}_p(R^d)$ and $\nu \in \mathbf{P}_p(R^d)$ is defined as:
\begin{equation}
    SW_p^p(\mu, \nu) = \mathbb{E}_{\theta\sim U (\mathbb{S}^{d-1})}[W_p^p(\theta\sharp\mu, \theta\sharp\nu)]
\end{equation}
$W_p^p(\theta\sharp\mu, \theta\sharp\nu)$ has a closed-form solution $\int_0^1|F_{\theta\sharp\mu}^{-1}(z) - F_{\theta\sharp\nu}^{-1}(z)|^pdz$ with $F^{-1}$ denotes the inverse CDF function. For discrete measures $\mu$ and $\nu$ with at most $m$ supports, the time and memory complexity of the Monte Carlo estimation of SWD are $O(Lm\log m)$ and $O(Lm)$ respectively. Hence, SWD is naturally suitable for comparing large-scale meshes. We do this by replace the expectation by the average from $\theta_1, \theta_2, ..., \theta_L$ that are drawn i.i.d from $U(\mathbb{S}^{d-1})$, where $L$ is the number of projections. So we have:
\begin{equation}
    \hat{SW}^p_p = \frac{1}{L}\sum_{l=1}^L W_p^p(\theta_l\sharp\mu, \theta_l\sharp\nu)
\end{equation}
In our implementation, we always use $p=2, L=100$.

\begin{figure*}[htbp]
\centering
\begin{minipage}[t]{0.48\textwidth}
\centering
\includegraphics[scale=0.162]{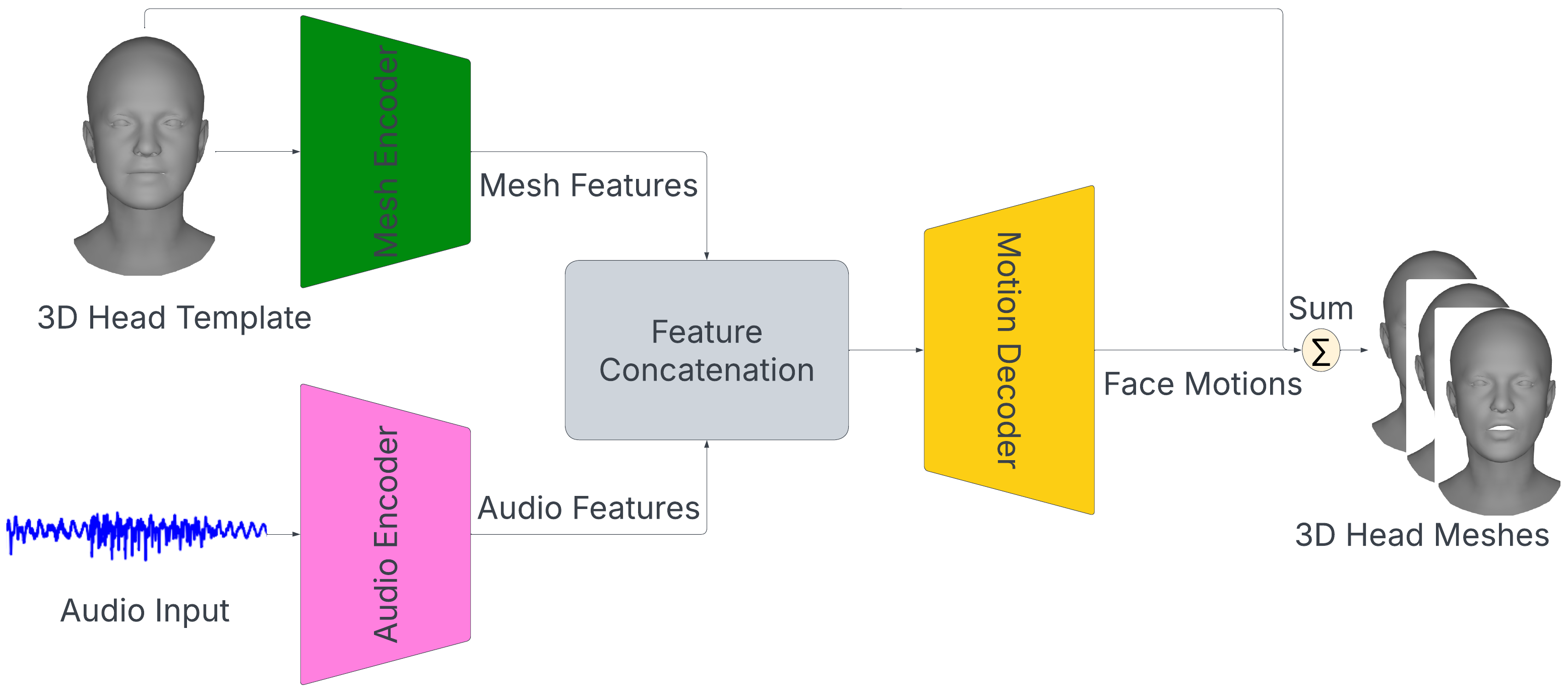}
\caption{System Pipeline}
\end{minipage}
\hfill
    \begin{tikzpicture}
        \draw[dashed, thick] (0,0) -- (0,3.5); 
    \end{tikzpicture}
    \hfill
\begin{minipage}[t]{0.48\textwidth}
\centering
\includegraphics[scale=0.253]{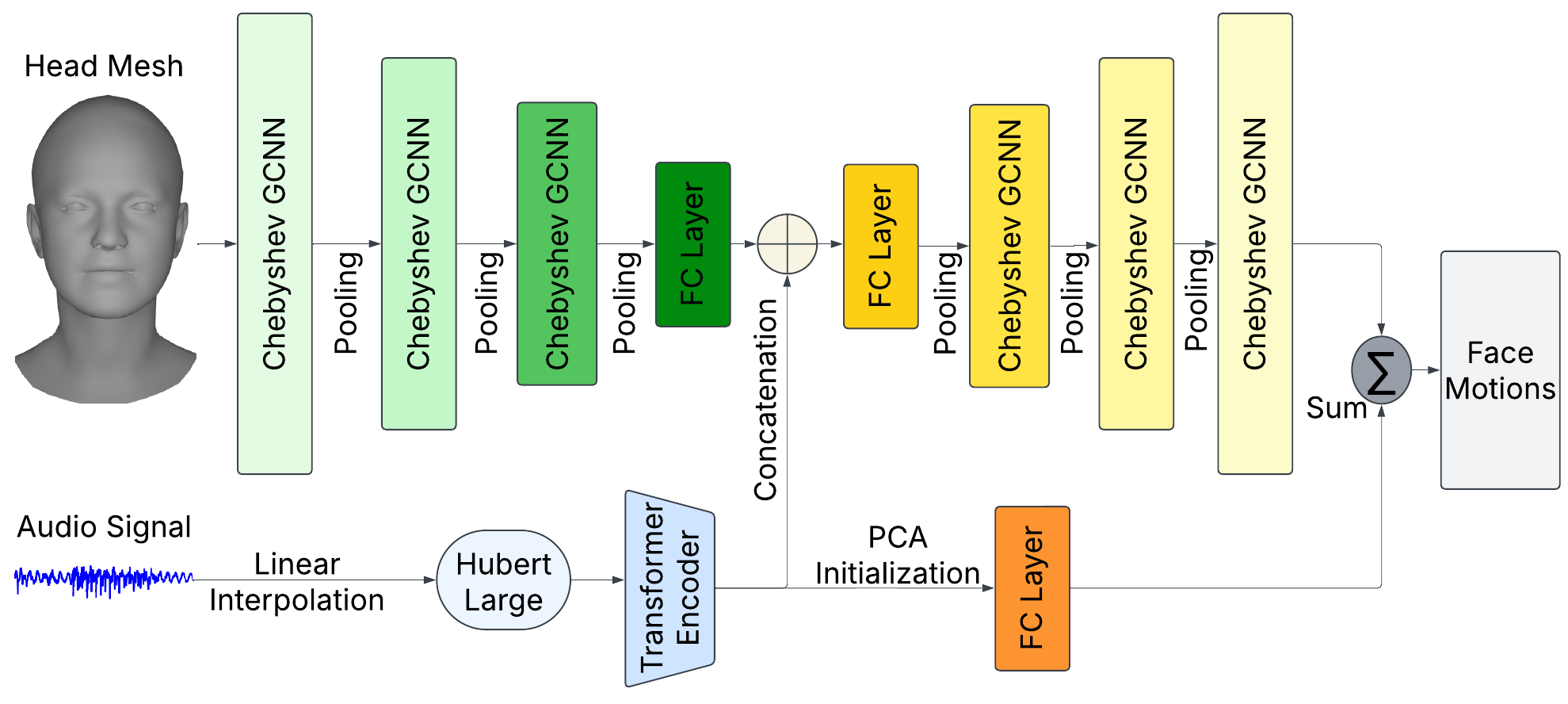}
\caption{Network Architecture}
\end{minipage}
\caption*{We use an auto-encoder structure in Figure 1 that takes an audio clip and a 3D head template as input. Mesh features and audio features are extracted by corresponding encoders and concatenated for a face motion decoder. A Head mesh sequence can be generated from the face motions at each frame. Detailed modules of the neural network are shown in Figure 2.}
\end{figure*}

\subsection{Pipeline and Network Architecture}

The network architecture is illustrated in Figure 2 and follows an auto-encoder pipeline in Figure 1 (details of network layers can be referred from supplementary material). It features two encoders: one for the template mesh and one for the audio signal. The mesh encoder extracts features from the template mesh, and the audio encoder processes the audio input. The extracted features are concatenated and fed into a motion decoder, which computes vertex displacements. The final mesh is generated by applying these displacements to the template mesh.
The mesh encoder contains 3 Chebyshev convolution layers, 3 pooling layers and 1 fully connected layer. Each Chebyshev convolution layer is followed by a pooling layer to downsample the mesh to its 1/4 size, i.e., $M_i = M_{i-1}/4$. E.g., For Multiface dataset, the mesh series are in size of $\{7306, 1827, 457, 115\}$. For VOCASET, the mesh sizes are $\{5023, 1256, 314, 79\}$. 
For audio preprocessing, we first use the pre-trained Huber model (Hubert Large), and linearly interpolate the output to an array with length of $T$ where \(T\) is the length of the audio clip. The audio extractor then conducts temporal convolution and uses a transformer encoder to convert the audio features into contextualized speech representations.
We compute the mesh displacements in 2 parts. First we try to exploit the linearity of face motions which is commonly used in 3DMM techniques \cite{egger20203d}. We use a fully-connected layer initialized by performing principle component analysis (PCA) on the training set to compute 50 linear face motion components from the audio latent code. Then we refine the face motions by adding the output of the motion decoder which uses a very similar architecture as the motion encoder. Once we get the facial motions for each frame, we can compute the mesh sequences of a 3D talking head.

\subsection{Implementation \& Training Details}

The implementation utilizes PyTorch and PyTorch Geometric. Given that each dataset has a different input mesh shape, models are implemented and trained separately for each dataset, but with consistent settings. The batch size is set to 32. We use the AdamW optimizer with a weight decay of \(0.01\) and a learning rate of \(1 \times 10^{-4}\). The training objective consists of three components. The first component is to minimize the reconstruction error per frame:
\begin{equation}
    L_r = \sum_{j=1}^{T}\sum_{i=1}^{N}||\hat{y}_i^j - y_i^j||_2
\end{equation}
where $T$ is the number of frames and $N$ is the number of vertices of a mesh.
The second component is to minimize the difference between consecutive frames for both the predicted outputs and the training vertices:
\begin{equation}
    L_v = \sum_{j=1}^{T-1}\sum_{i=1}^{N}||(\hat{y}_i^{j+1} - \hat{y}_i^j) - (y_i^{j+1} - y_i^j)||_2
\end{equation}
The third component is the SWD computed from every predicted mesh and target mesh for every frame. 
\begin{equation}
    L_{SW} = \sum_{j=1}^{T}\frac{1}{L}\sum_{l}^{L}W_p^p(\theta_l\sharp\tilde{\mu}^{\hat{M_j}}, \theta_l\sharp\tilde{\mu}^{Mj})
\end{equation}
The overall training objective is to minimize the toal loss with sparsity regularization which minimizes the $l2$ norm of $W$:
\begin{equation}
    L = L_r + \beta_1 L_v + \beta_2 L_{SW} + \beta_3 ||W||_2
\end{equation}
where $\beta_1=10, \beta_2 = 1, \beta_3=0.01$ in our implementation. We choose AdamW optimizer with a learning rate of $1e-4$.

Training is conducted on a single NVIDIA RTX 5000 GPU. Each epoch takes approximately 2.5 minutes. 100 epochs (we pick the best parameter weights by validation set) are required for convergence. The inference time per frame is approximately 0.007 seconds, making the model suitable for real-time applications.

\section{Experiment}

\begin{table*}[!htbp]
    \centering
    \begin{tabular}{c||c c c c c|| c c c c c}
        \hline
        Methods& \multicolumn{5}{|c||}{VOCASET Dataset (mm)} & \multicolumn{4}{|c}{Multiface Dataset (mm)} \\
        \hline
        \hline
         & $E_{max}^{lip}$ & $E_{mean}^{lip}$ & $E_{mean}^{face}$ & $E_{mean}^{head}$ &$DTW_{lip}$ &$E_{max}^{lip}$ & $E_{mean}^{lip}$ & $E_{mean}^{face}$ & $E_{mean}^{head}$ &$DTW_{lip}$\\
        \hline
        VOCA & 5.11 & 2.93 & 1.31 &1.00 &740.05 & 11.60 & 2.19 & 1.30 & 1.55 & 707.38\\
        \hline
        FaceFormer & 8.95 & 5.72 & 2.98 & 2.95 & 909.72 & 11.68 & 2.17 & 1.28 & 1.54 & 801.64\\
        \hline
        CodeTalker & 8.90 & 5.77 & 2.99 & 2.93& 912.50 &12.08 & 3.12 & 1.72 & 1.90 &889.52\\
        \hline
        \hline
        \textbf{OT-Talk} & \textbf{5.01} & \textbf{2.81} & \textbf{1.26} & \textbf{0.87} &\textbf{710.48} & \textbf{10.17} & \textbf{1.86} & \textbf{1.15} & \textbf{1.40} & \textbf{616.45}\\
        \hline
    \end{tabular}
    \caption{Quantitative Results. Lower Means Better for All Metrics}
\end{table*}

\textbf{Datasets.} We use two public datasets, VOCASET and Multiface. BiWi is also a popular dataset but no longer available, so it is not included in this experiment. VOCASET comprises 480 facial motion sequences from 12 subjects, with 40 sequences per subject. Each sequence is recorded at 60 FPS and lasts 3-4 seconds, and the 3D head mesh has 5023 vertices. To ensure fair comparison, we follow the same split as VOCA: 8 subjects for training, 2 for validation, and 2 for testing, with no overlap of sentences or subjects. The Multiface dataset includes audio-mesh pairs from 13 subjects, captured at 30 FPS. One subject provides 12 sentences, while the others provide 50 sentences, all with the same content. We use a similar split: 8 subjects for training, 3 for validation, and 2 for testing. We allocate 30 sentences for training, 10 for validation, and 10 for testing. The training, validation, and test sets are completely disjoint.

\textbf{Baseline Methods.} We compare our method with three state-of-the-art techniques: FaceFormer \cite{fan2022faceformer}, CodeTalker \cite{xing2023codetalker} and VOCA \cite{cudeiro2019capture}. All these methods use an identity code and speech input. Since the subjects in the validation and test sets are unseen during training, we generate results for all identity conditions and select the \textbf{best}-performing one. The official codebases for FaceFormer, CodeTalker and VOCA provide pre-trained weights on VOCASET, which we use directly. FaceFormer, CodeTalker and VOCA produce results for VOCASET at 30 FPS, so we downsample the ground truth frames by half for evaluation. For the Multiface dataset, we adhere to the official implementations for both training and testing. Additional details on implementation and results with different identity conditions are available in the supplementary materials.

\subsection{Quantitative Results}
We conduct a quantitative evaluation across several metrics. The maximal lip vertex error, $E_{max}^{lip}$, computes the maximum Euclidean distance between vertices in the lip region of the synthesized ground truth head meshes and averages this error across frames. The mean lip vertex error, $E_{mean}^{lip}$, calculates the average Euclidean distance between lip vertices. Besides assessing lip movement synchronization, we also evaluate the reconstruction accuracy of the inner face region ($E_{mean}^{face}$) and the full head ($E_{mean}^{head}$) by computing the average Euclidean distance. For the VOCASET dataset, we use facial part masks provided by the FLAME \cite{li2017learning} model, whereas for the Multiface dataset, we manually capture the lip and face masks from a template mesh using MeshLab \cite{LocalChapterEvents:ItalChap:ItalianChapConf2008:129-136}.
VOCA, FaceFormer, and CodeTalker rely on one-hot labels for identity conditioning. We compute results for all 8 identities and select the optimal case for comparison. Identity conditioning largely impacts the generated results. Hence, to demonstrate the superiority of our approach, we generate results under all identities for the baselines and list their minimum quantitative errors for comparison  (results for other identities are provided in the supplementary materials). Our approach does not use one-hot label but learn identity condition from mesh geometry. The quantitative results are summarized in Table 1. 

Our method achieves the best score on all metrics for both datasets. On Multiface dataset, our approach show a more significant gap against the sub-optimal approach. Though VOCASET and Multiface are all 4D scans of head meshes, they have obvious differences which affects the experiment results. VOCASET dataset is smoothed by filtering eye movements and extreme expressions, while Multiface keeps these face dynamics like blinking eyes (ChebNet used in our approach is more suitable in this case given its representation ability for 3D faces). Multiface dataset is also more noisy than VOCASET which contains noncontinuous changes between consecutive frames and our method turns out to be more robust in this case. VOCA achieves the suboptimal quantitative results on VOCASET largely because the VOCA's face motion decoder exploits the linearity of face movements, which is more suitable for VOCASET using the Flame model as mesh template. Additionally, its last linear layer is initialized using PCA components derived from vertex displacements (also adopted by our approach), giving it prior knowledge about the face motion space. Although FaceFormer and CodeTalker obviously have larger estimation error on face motions, they still have relatively close qualitative performance compared to other methods. 
For Multiface dataset, FaceFormer has suboptimal quantitative error, but it failed to synthesize lip motions for many cases due to over-smooth problem (shown in qualitative evaluation and user perception study). VOCA can only drive lip motions while it cannot synthesize eye movements or other face emotions. CodeTalker turns out to be inaccurate on quantitative evaluation, but can produce extreme facial expressions (sometimes even better than our approach, though failed to synchronize with audio for many cases).   

\begin{table*}[!htbp]
    \centering
    \begin{tabular}{c||c c c c c|| c c c c c}
        \hline
        Methods& \multicolumn{5}{|c||}{VOCASET Dataset (mm)} & \multicolumn{4}{|c}{Multiface Dataset (mm)} \\
        \hline
        \hline
         & $E_{max}^{lip}$ & $E_{mean}^{lip}$ & $E_{mean}^{face}$ & $E_{mean}^{head}$ &$DTW_{lip}$ &$E_{max}^{lip}$ & $E_{mean}^{lip}$ & $E_{mean}^{face}$ & $E_{mean}^{head}$ &$DTW_{lip}$\\
        \hline
        \textbf{OT-Talk} & \textbf{5.01} & \textbf{2.81} & \underline{1.26} & \underline{0.87} &\textbf{710.48} & \textbf{10.17} & \textbf{1.86} & \textbf{1.15} & \textbf{1.40} & \textbf{616.45}\\
        \hline
        w/o OT & 5.25 & 3.00 & 1.30 & 0.87 &749.58 & 10.58 & 1.92 & 1.19 & 1.45 & 635.02\\
        \hline
        w/o Transformer & 5.27 & 3.04 & \textbf{1.19} & \textbf{0.86} & 909.22 & 11.23 & 2.11 & 1.25 & 1.51 & 674.10\\
        \hline
        w/o ChebNet & 8.39 & 5.29 & 4.48 & 4.66 & 1314.48 & 14.07 & 5.11 & 4.89 & 6.92 & 1649.44\\
        \hline
    \end{tabular}
    \caption{Ablation Study to investigate the impact of individual components in OT-Talk. Lower Means Better for All Metrics}
\end{table*}

\subsection{Ablation Study}

We also do ablation evaluation on our learning model by removing different modules including the OT loss (only keep reconstruction and velocity loss), the transformer audio encoder (only keep temporal convolution), and ChebNet (decode face motions from audio decoder only). The quantitative results are listed in Table 2. ChebNet has the greatest impact on performance because without it, the neural network does not have enough representation space for head meshes or face motions. Other components also help to optimize our learning model, especially in learning lip motions.

\begin{figure}
    \centering
    {re\red{war}d}
    \begin{minipage}{.47\textwidth}
        \centering
        \includegraphics[width=0.193\textwidth]{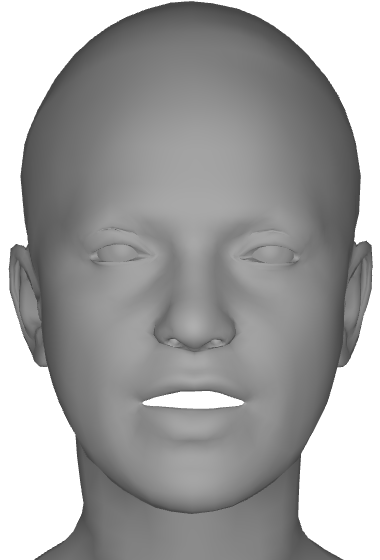}
        \includegraphics[width=0.193\textwidth]{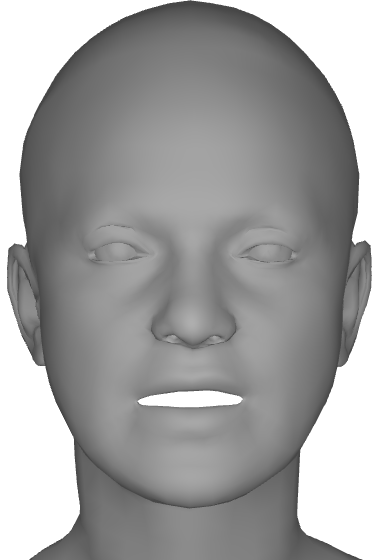}
        \includegraphics[width=0.193\textwidth]{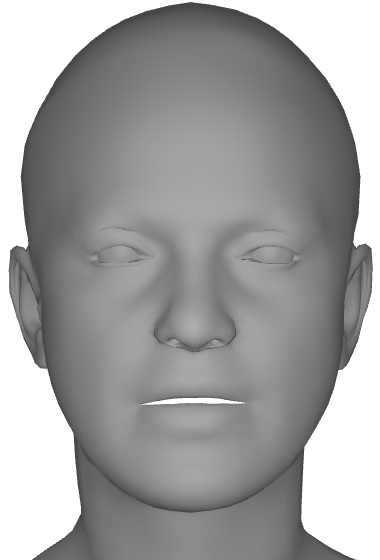}
        \includegraphics[width=0.193\textwidth]{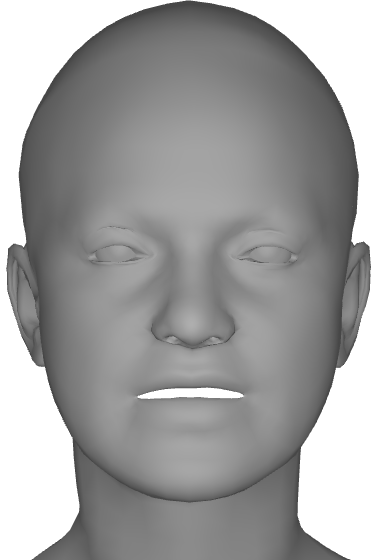}
        \includegraphics[width=0.193\textwidth]{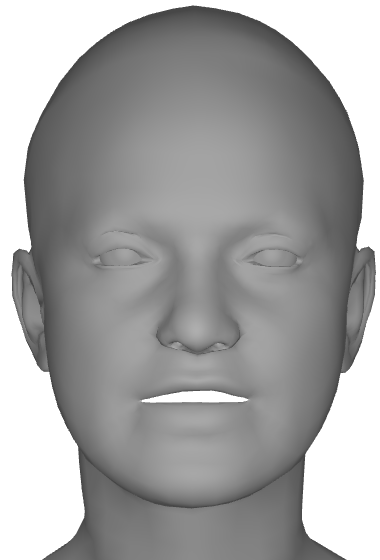}
    \end{minipage}
    \begin{minipage}{.47\textwidth}
        \centering
        \includegraphics[width=0.193\textwidth]{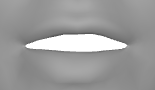}
        \includegraphics[width=0.193\textwidth]{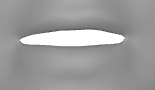}
        \includegraphics[width=0.193\textwidth]{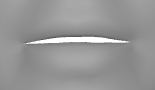}
        \includegraphics[width=0.193\textwidth]{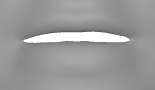}
        \includegraphics[width=0.193\textwidth]{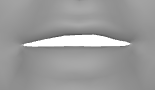}
    \end{minipage}
\vspace{.3cm}

    \centering
    {\red{hy}dride}
    \begin{minipage}{.47\textwidth}
        \centering
        \includegraphics[width=0.193\textwidth]{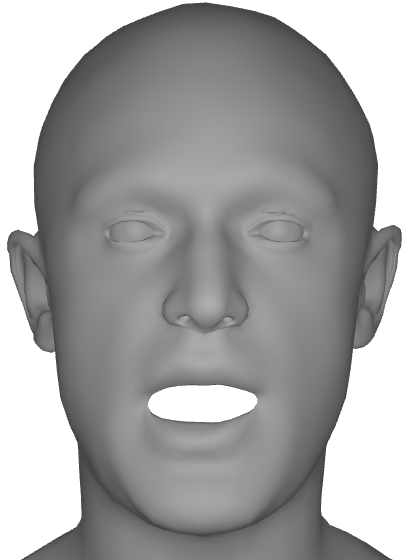}
        \includegraphics[width=0.193\textwidth]{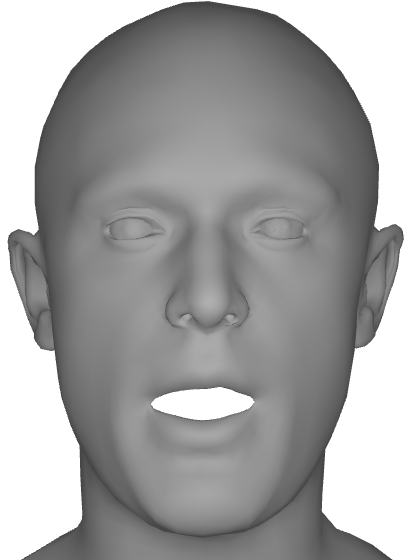}
        \includegraphics[width=0.193\textwidth]{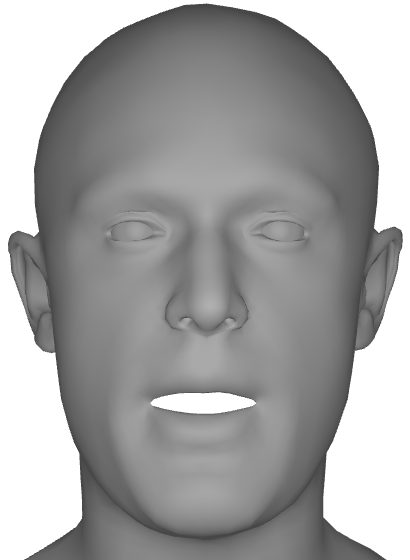}
        \includegraphics[width=0.193\textwidth]{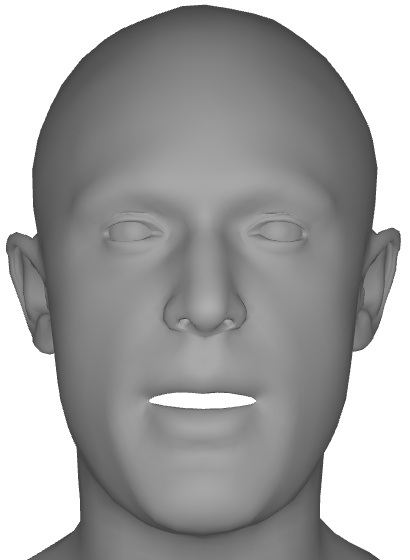}
        \includegraphics[width=0.193\textwidth]{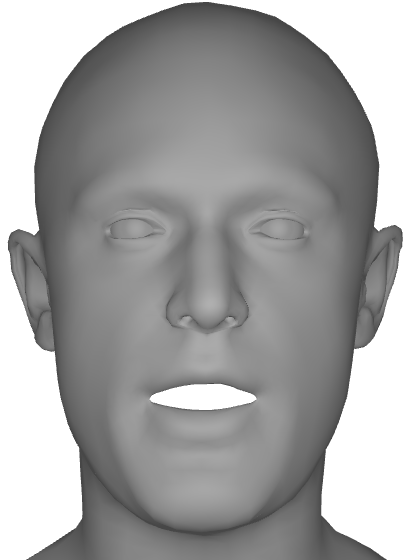}
    \end{minipage}
    \begin{minipage}{.47\textwidth}
        \centering
        \includegraphics[width=0.193\textwidth]{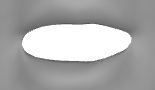}
        \includegraphics[width=0.193\textwidth]{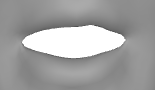}
        \includegraphics[width=0.193\textwidth]{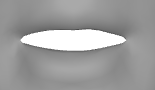}
        \includegraphics[width=0.193\textwidth]{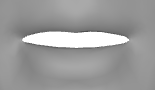}
        \includegraphics[width=0.193\textwidth]{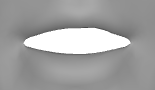}
    \end{minipage}
\vspace{.3cm}

    \centering
    {pl\red{an}}
    \begin{minipage}{.47\textwidth}
        \centering
        \includegraphics[width=0.193\textwidth]{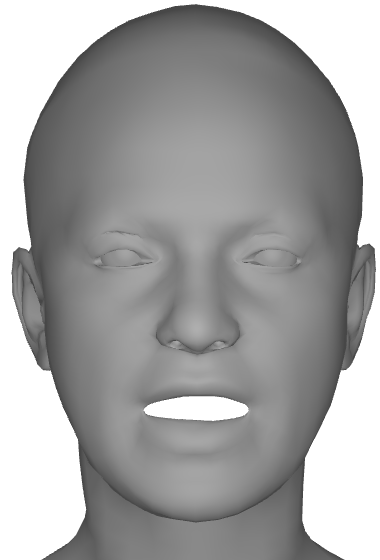}
        \includegraphics[width=0.193\textwidth]{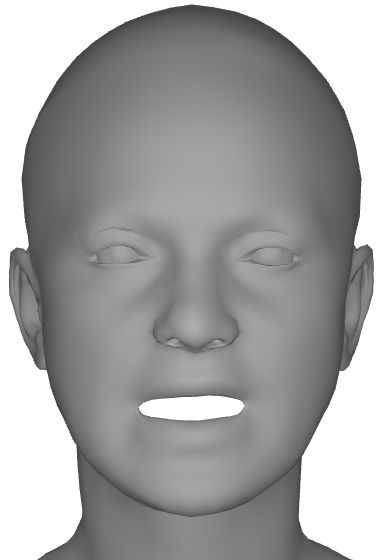}
        \includegraphics[width=0.193\textwidth]{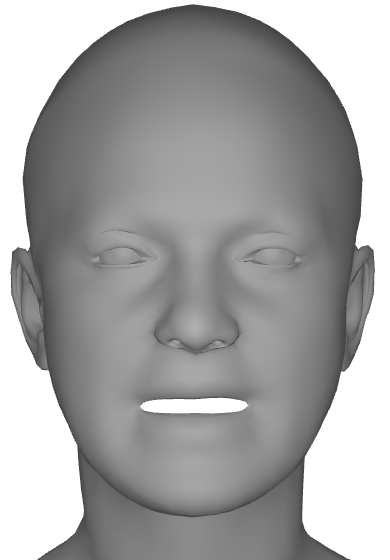}
        \includegraphics[width=0.193\textwidth]{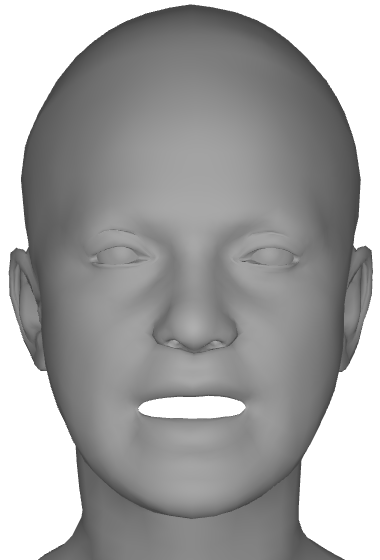}
        \includegraphics[width=0.193\textwidth]{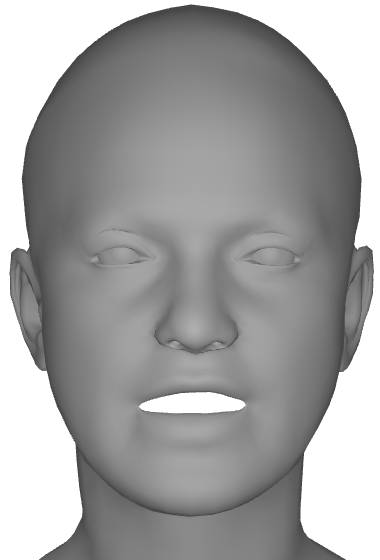}
    \end{minipage}
    \begin{minipage}{.47\textwidth}
        \centering
        \includegraphics[width=0.193\textwidth]{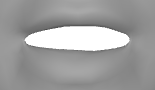}
        \includegraphics[width=0.193\textwidth]{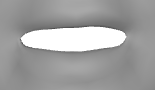}
        \includegraphics[width=0.193\textwidth]{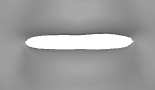}
        \includegraphics[width=0.193\textwidth]{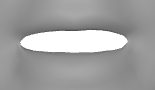}
        \includegraphics[width=0.193\textwidth]{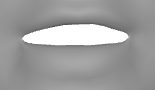}
    \end{minipage}
\vspace{.3cm}

    \centering
    {wh\red{a}t}
    \begin{minipage}{.47\textwidth}
        \centering
        \includegraphics[width=0.193\textwidth]{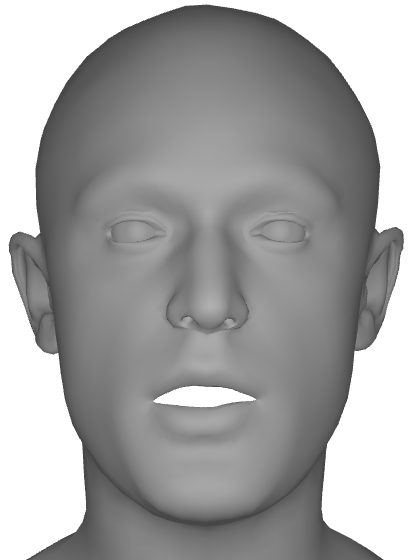}
        \includegraphics[width=0.193\textwidth]{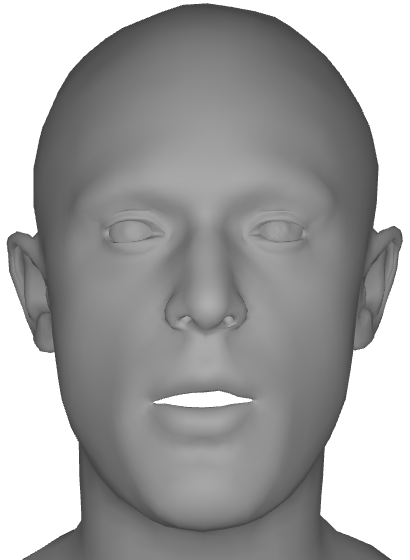}
        \includegraphics[width=0.193\textwidth]{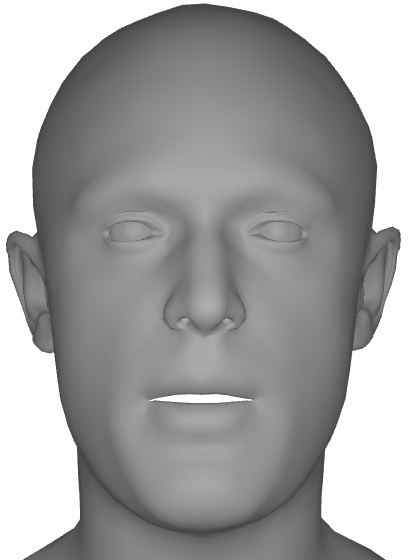}
        \includegraphics[width=0.193\textwidth]{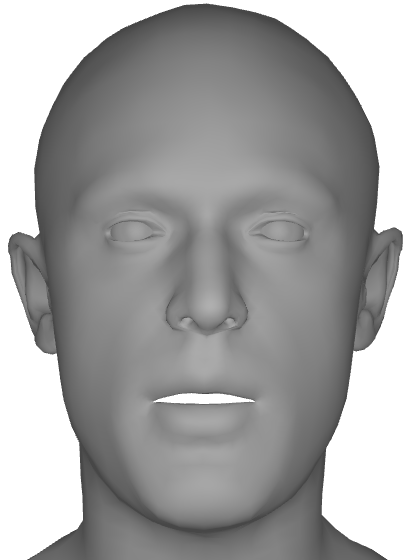}
        \includegraphics[width=0.193\textwidth]{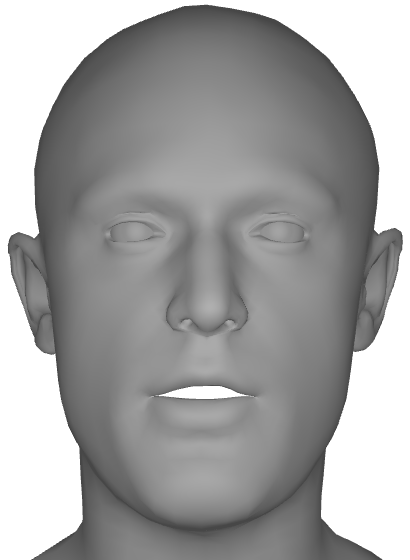}
    \end{minipage}
    \begin{minipage}{.47\textwidth}
        \centering
        \includegraphics[width=0.193\textwidth]{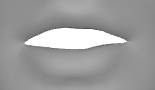}
        \includegraphics[width=0.193\textwidth]{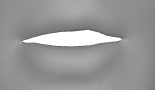}
        \includegraphics[width=0.193\textwidth]{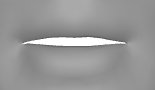}
        \includegraphics[width=0.193\textwidth]{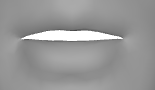}
        \includegraphics[width=0.193\textwidth]{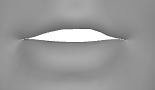}
    \end{minipage}
\vspace{-.4cm}

    \caption*{\ \ \ \ \ GT \ \  \ \ \ \ \  \ \  OT-Talk\ \ \ \ \  FaceFormer\   \ CodeTalker\ \ \ VOCA}
    \caption{VOCASET Qualitative results (face/lip motions to pronounce words). OT-Talk achieves the closest lip shapes with the ground truth (GT). For word "hydride", the subject has a drastic lip motion and all the methods cannot synthesize such shape.}
\end{figure}

\begin{figure}
    \centering
    {c\red{a}tch}
    \begin{minipage}{.47\textwidth}
        \centering
        \includegraphics[width=0.19\textwidth]{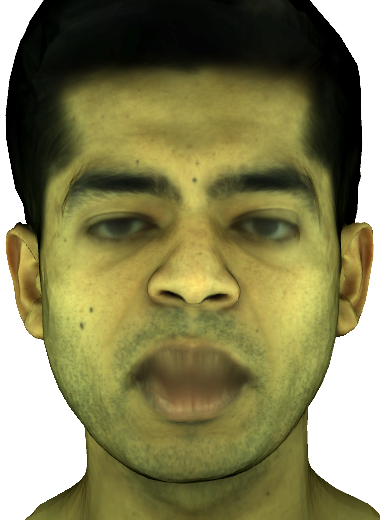}
        \includegraphics[width=0.19\textwidth]{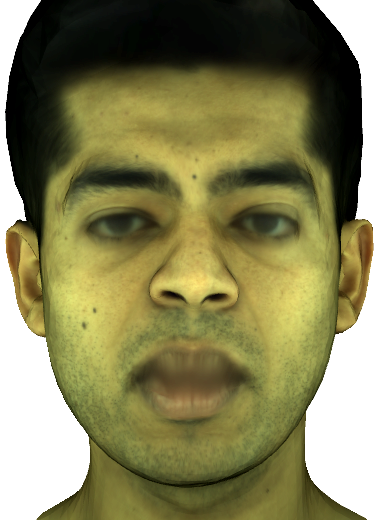}
        \includegraphics[width=0.19\textwidth]{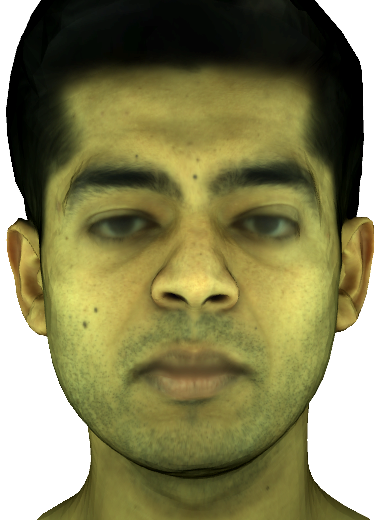}
        \includegraphics[width=0.19\textwidth]{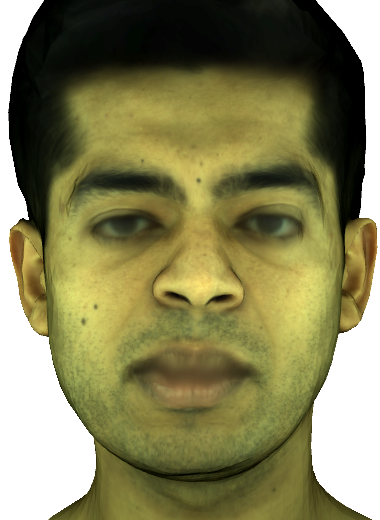}
        \includegraphics[width=0.19\textwidth]{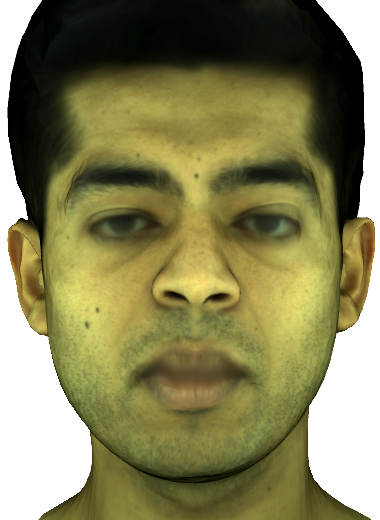}
    \end{minipage}
    \begin{minipage}{.47\textwidth}
        \centering
        \includegraphics[width=0.19\textwidth]{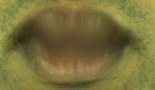}
        \includegraphics[width=0.19\textwidth]{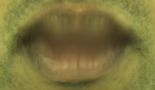}
        \includegraphics[width=0.19\textwidth]{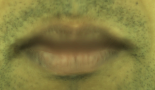}
        \includegraphics[width=0.19\textwidth]{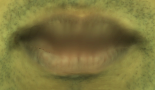}
        \includegraphics[width=0.19\textwidth]{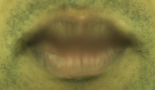}
    \end{minipage}
    
\vspace{.4cm}

    \centering
    {cl\red{a}ss}
    \begin{minipage}{.47\textwidth}
        \centering
        \includegraphics[width=0.19\textwidth]{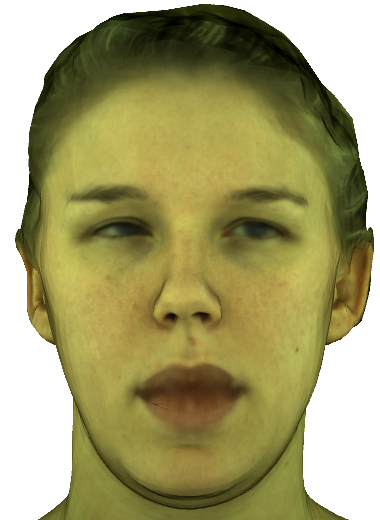}
        \includegraphics[width=0.19\textwidth]{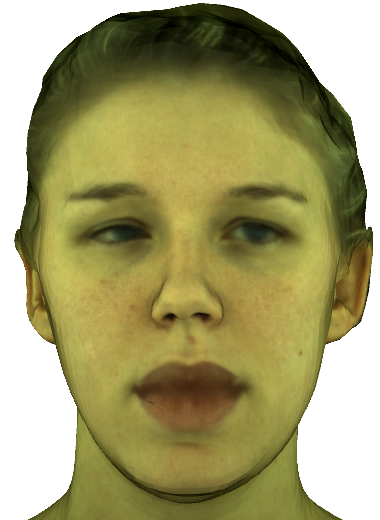}
        \includegraphics[width=0.19\textwidth]{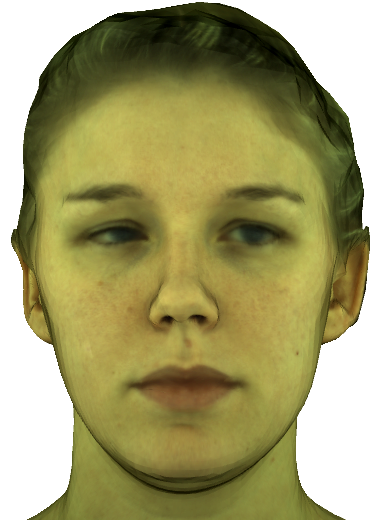}
        \includegraphics[width=0.19\textwidth]{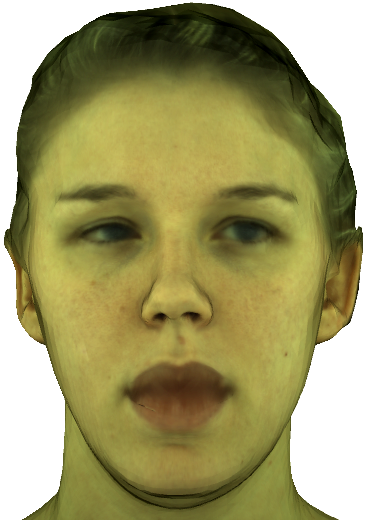}
        \includegraphics[width=0.19\textwidth]{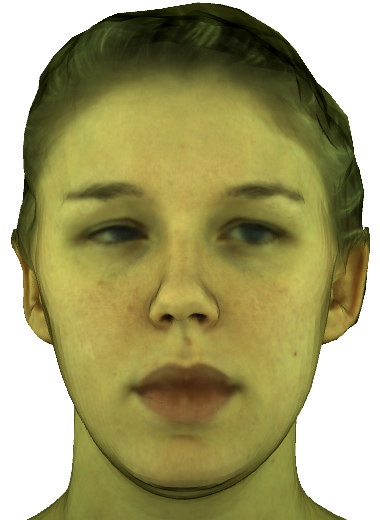}
    \end{minipage}
    \begin{minipage}{.47\textwidth}
        \centering
        \includegraphics[width=0.19\textwidth]{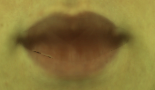}
        \includegraphics[width=0.19\textwidth]{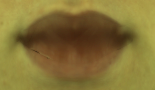}
        \includegraphics[width=0.19\textwidth]{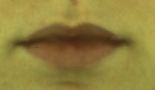}
        \includegraphics[width=0.19\textwidth]{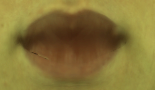}
        \includegraphics[width=0.19\textwidth]{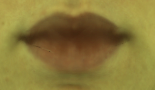}
    \end{minipage}
\vspace{.4cm}

    \centering
    {h\red{ow}}
    \begin{minipage}{.47\textwidth}
        \centering
        \includegraphics[width=0.19\textwidth]{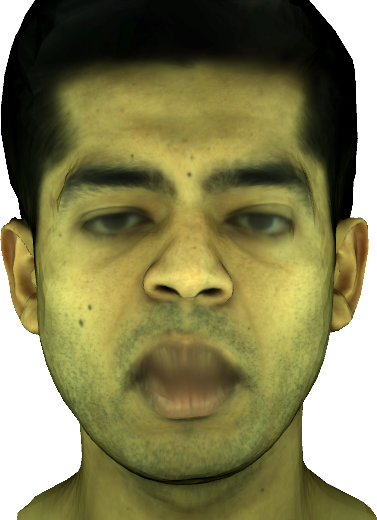}
        \includegraphics[width=0.19\textwidth]{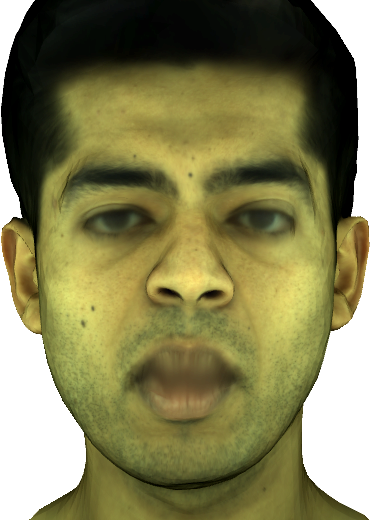}
        \includegraphics[width=0.19\textwidth]{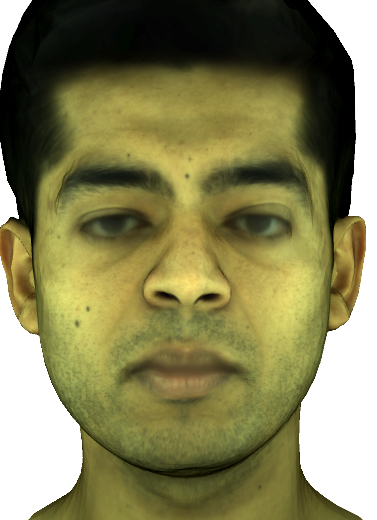}
        \includegraphics[width=0.19\textwidth]{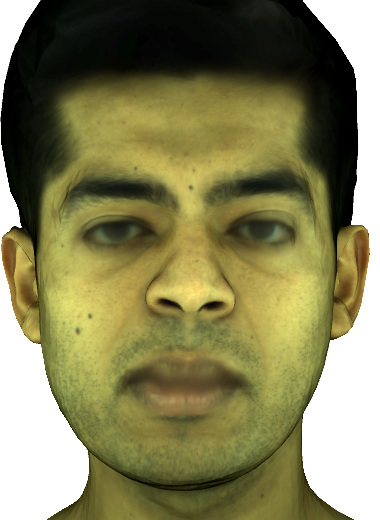}
        \includegraphics[width=0.19\textwidth]{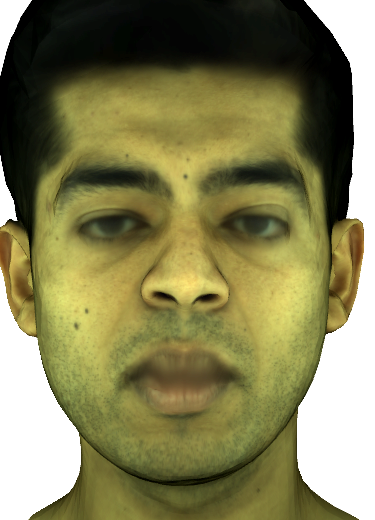}
    \end{minipage}
    \begin{minipage}{.47\textwidth}
        \centering
        \includegraphics[width=0.19\textwidth]{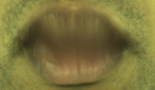}
        \includegraphics[width=0.19\textwidth]{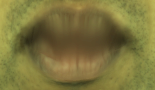}
        \includegraphics[width=0.19\textwidth]{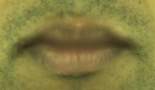}
        \includegraphics[width=0.19\textwidth]{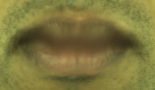}
        \includegraphics[width=0.19\textwidth]{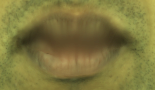}
    \end{minipage}
\vspace{.4cm}

    \centering
    {b\red{uy}}
    \begin{minipage}{.47\textwidth}
        \centering
        \includegraphics[width=0.19\textwidth]{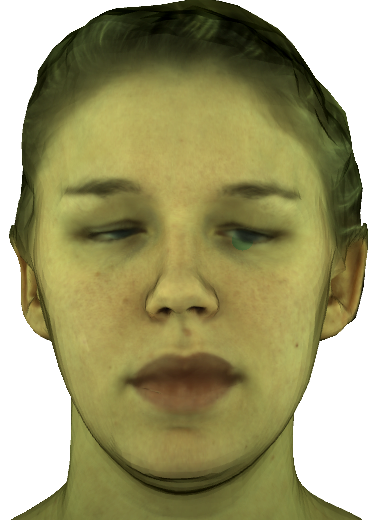}
        \includegraphics[width=0.19\textwidth]{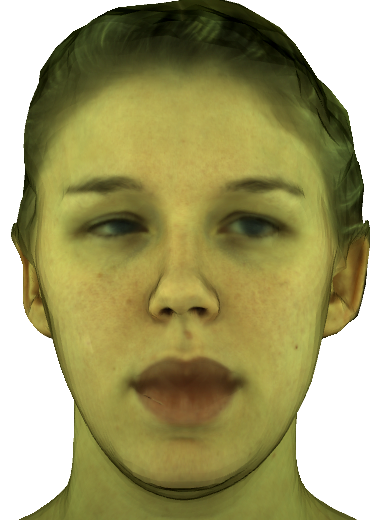}
        \includegraphics[width=0.19\textwidth]{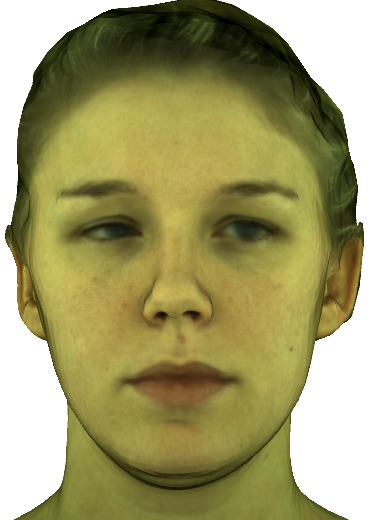}
        \includegraphics[width=0.19\textwidth]{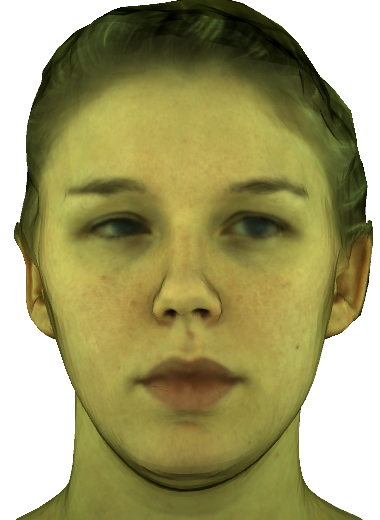}
        \includegraphics[width=0.19\textwidth]{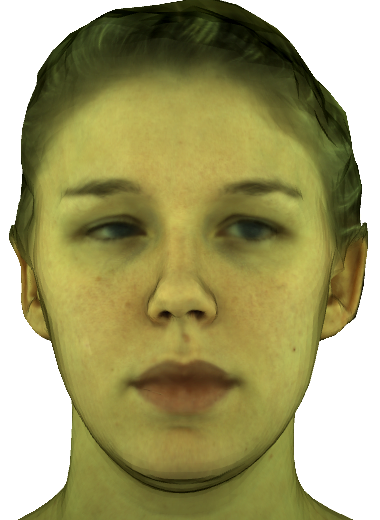}
    \end{minipage}
    \begin{minipage}{.47\textwidth}
        \centering
        \includegraphics[width=0.19\textwidth]{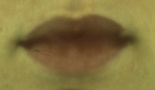}
        \includegraphics[width=0.19\textwidth]{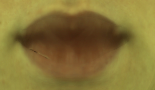}
        \includegraphics[width=0.19\textwidth]{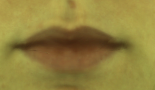}
        \includegraphics[width=0.19\textwidth]{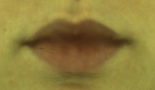}
        \includegraphics[width=0.19\textwidth]{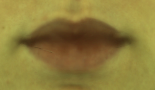}
    \end{minipage}
    
    \caption*{\ \ \ GT \ \  \ \ \ \ \  \ \  OT-Talk\ \ \ \ \  FaceFormer\   \ \ CodeTalker\ \ VOCA}
    \caption{Multiface Qualitative results (face/lip motions to pronounce words). OT-Talk is the closest to ground truth (GT) on lip shapes. In the last row, ground truth shows an eye motion but all methods cannot synthesize it.}
\end{figure}

\begin{figure}[!htbp]
    
    \begin{subfigure}{.45\textwidth}
        \centering
        \includegraphics[width=1\linewidth]{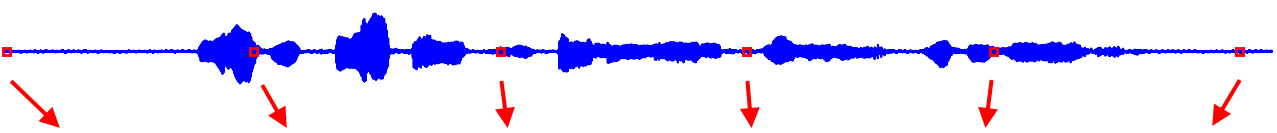}
    \end{subfigure}
    \vspace{-.4cm}
    \caption*{\textbf{Ground Truth}}
    
    \centering
    \begin{subfigure}{0.075\textwidth}
        \centering
        \includegraphics[width=\linewidth]{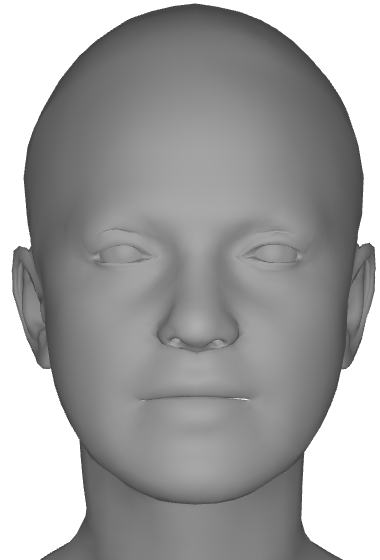}
    \end{subfigure}
    \hfill
    \begin{subfigure}{0.075\textwidth}
        \centering
        \includegraphics[width=\linewidth]{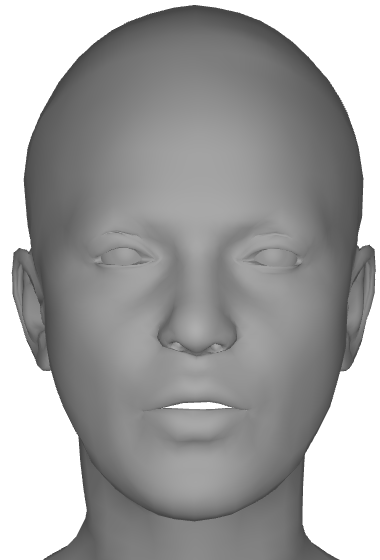}
    \end{subfigure}
    \hfill
    \begin{subfigure}{0.075\textwidth}
        \centering
        \includegraphics[width=\linewidth]{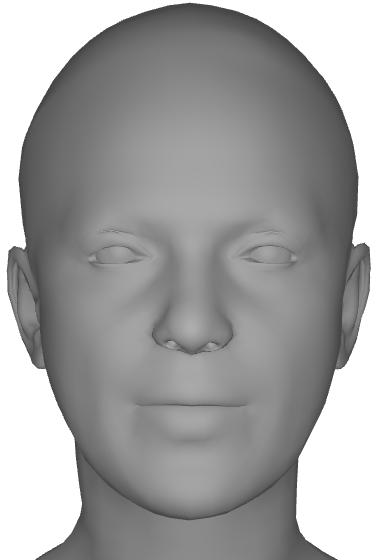}
    \end{subfigure}
    \hfill
    \begin{subfigure}{0.075\textwidth}
        \centering
        \includegraphics[width=\linewidth]{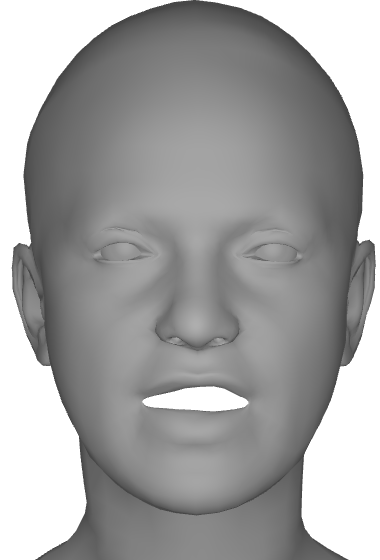}
    \end{subfigure}
    \hfill
    \begin{subfigure}{0.075\textwidth}
        \centering
        \includegraphics[width=\linewidth]{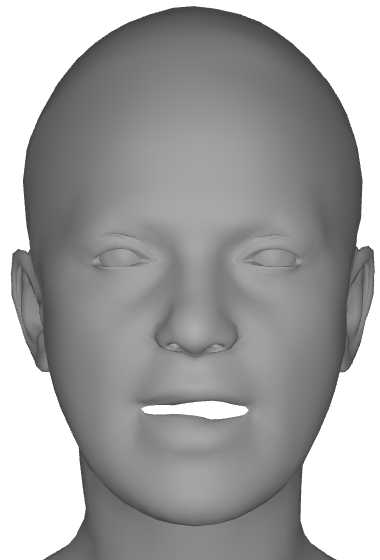}
    \end{subfigure}
    \hfill
    \begin{subfigure}{0.075\textwidth}
        \centering
        \includegraphics[width=\linewidth]{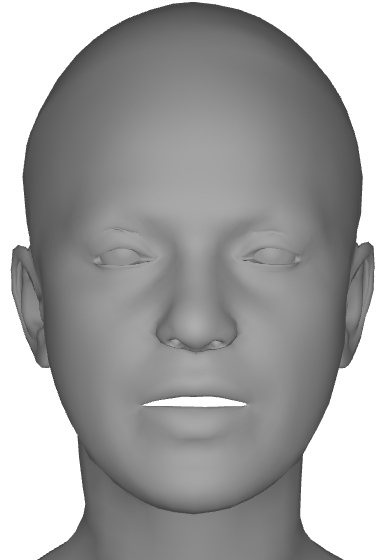}
    \end{subfigure}

    \centering
    \begin{subfigure}{0.075\textwidth}
        \centering
        \includegraphics[width=\linewidth]{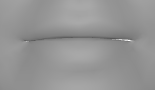}
    \end{subfigure}
    \hfill
    \begin{subfigure}{0.075\textwidth}
        \centering
        \includegraphics[width=\linewidth]{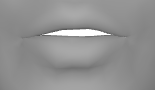}
    \end{subfigure}
    \hfill
    \begin{subfigure}{0.075\textwidth}
        \centering
        \includegraphics[width=\linewidth]{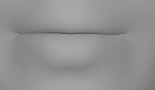}
    \end{subfigure}
    \hfill
    \begin{subfigure}{0.075\textwidth}
        \centering
        \includegraphics[width=\linewidth]{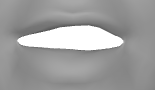}
    \end{subfigure}
    \hfill
    \begin{subfigure}{0.075\textwidth}
        \centering
        \includegraphics[width=\linewidth]{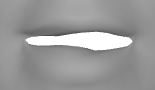}
    \end{subfigure}
    \hfill
    \begin{subfigure}{0.075\textwidth}
        \centering
        \includegraphics[width=\linewidth]{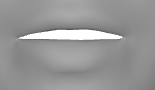}
    \end{subfigure}
    \vspace{-.2cm}
    \caption*{OT-Talk}

    \centering
    \begin{subfigure}{0.075\textwidth}
        \centering
        \includegraphics[width=\linewidth]{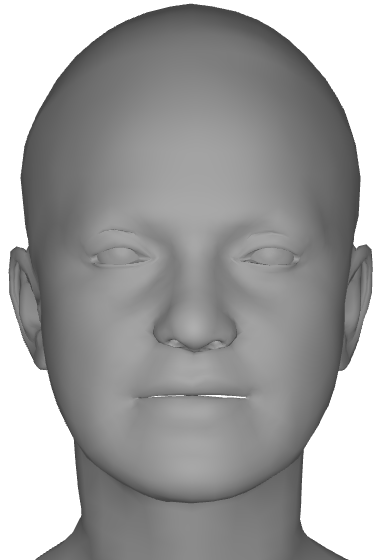}
    \end{subfigure}
    \hfill
    \begin{subfigure}{0.075\textwidth}
        \centering
        \includegraphics[width=\linewidth]{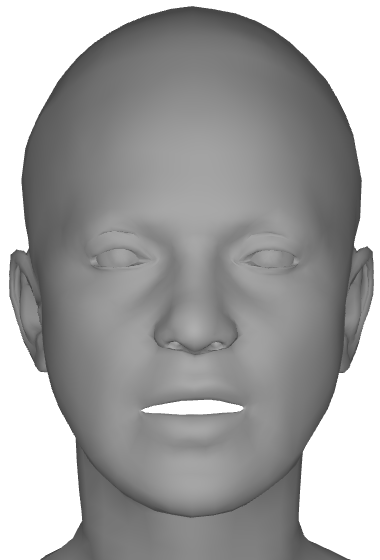}
    \end{subfigure}
    \hfill
    \begin{subfigure}{0.075\textwidth}
        \centering
        \includegraphics[width=\linewidth]{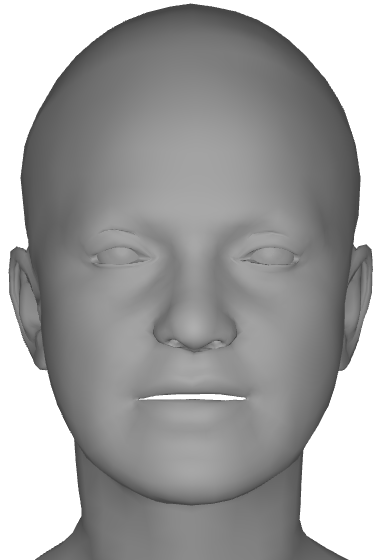}
    \end{subfigure}
    \hfill
    \begin{subfigure}{0.075\textwidth}
        \centering
        \includegraphics[width=\linewidth]{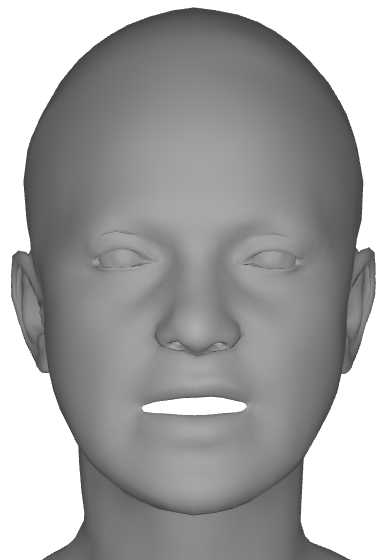}
    \end{subfigure}
    \hfill
    \begin{subfigure}{0.075\textwidth}
        \centering
        \includegraphics[width=\linewidth]{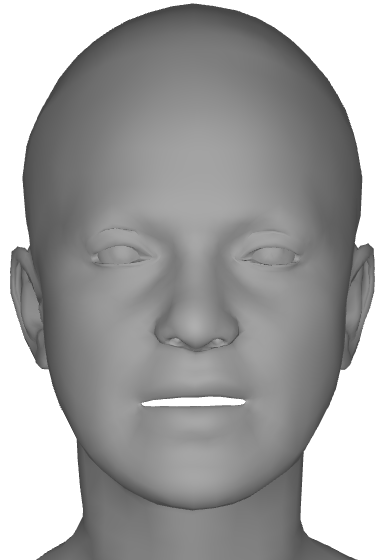}
    \end{subfigure}
    \hfill
    \begin{subfigure}{0.075\textwidth}
        \centering
        \includegraphics[width=\linewidth]{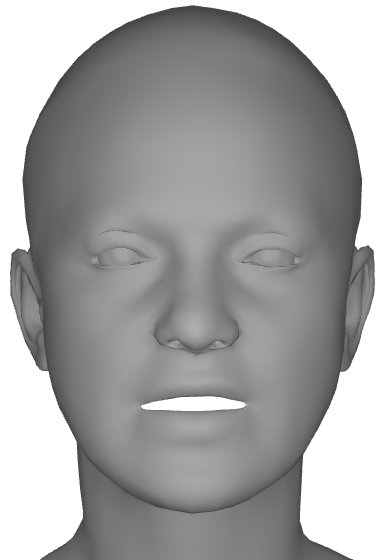}
    \end{subfigure}

    \centering
    \begin{subfigure}{0.075\textwidth}
        \centering
        \includegraphics[width=\linewidth]{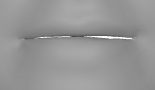}
    \end{subfigure}
    \hfill
    \begin{subfigure}{0.075\textwidth}
        \centering
        \includegraphics[width=\linewidth]{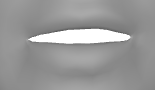}
    \end{subfigure}
    \hfill
    \begin{subfigure}{0.075\textwidth}
        \centering
        \includegraphics[width=\linewidth]{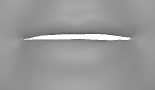}
    \end{subfigure}
    \hfill
    \begin{subfigure}{0.075\textwidth}
        \centering
        \includegraphics[width=\linewidth]{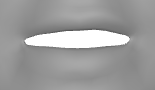}
    \end{subfigure}
    \hfill
    \begin{subfigure}{0.075\textwidth}
        \centering
        \includegraphics[width=\linewidth]{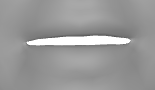}
    \end{subfigure}
    \hfill
    \begin{subfigure}{0.075\textwidth}
        \centering
        \includegraphics[width=\linewidth]{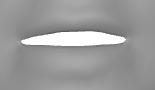}
    \end{subfigure}
    \vspace{-.2cm}
    \caption*{CodeTalker}

    \centering
    \begin{subfigure}{0.075\textwidth}
        \centering
        \includegraphics[width=\linewidth]{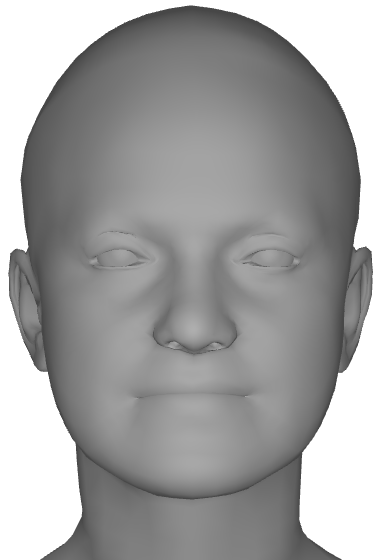}
    \end{subfigure}
    \hfill
    \begin{subfigure}{0.075\textwidth}
        \centering
        \includegraphics[width=\linewidth]{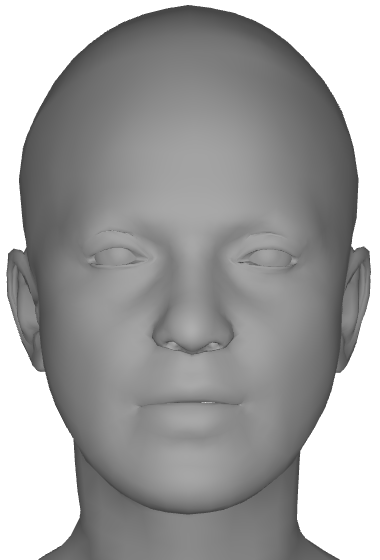}
    \end{subfigure}
    \hfill
    \begin{subfigure}{0.075\textwidth}
        \centering
        \includegraphics[width=\linewidth]{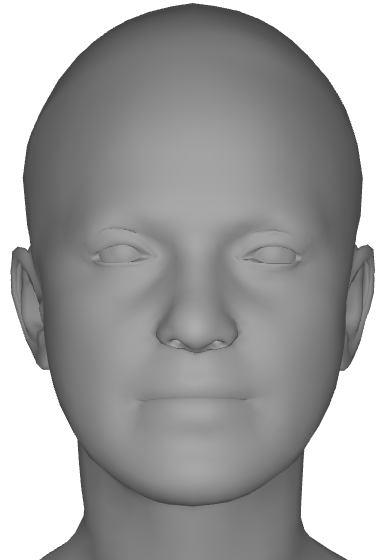}
    \end{subfigure}
    \hfill
    \begin{subfigure}{0.075\textwidth}
        \centering
        \includegraphics[width=\linewidth]{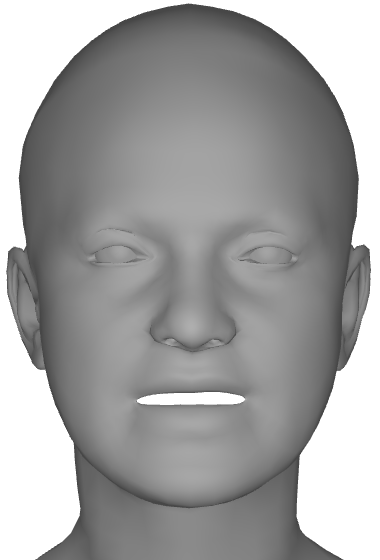}
    \end{subfigure}
    \hfill
    \begin{subfigure}{0.075\textwidth}
        \centering
        \includegraphics[width=\linewidth]{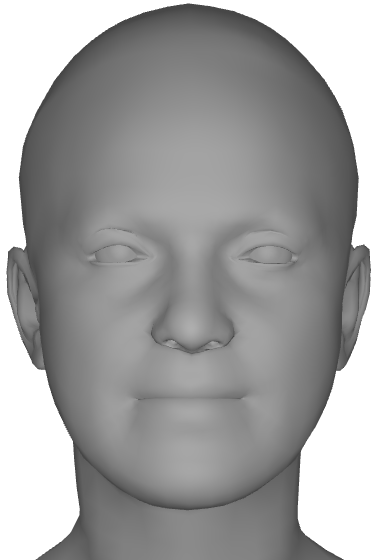}
    \end{subfigure}
    \hfill
    \begin{subfigure}{0.075\textwidth}
        \centering
        \includegraphics[width=\linewidth]{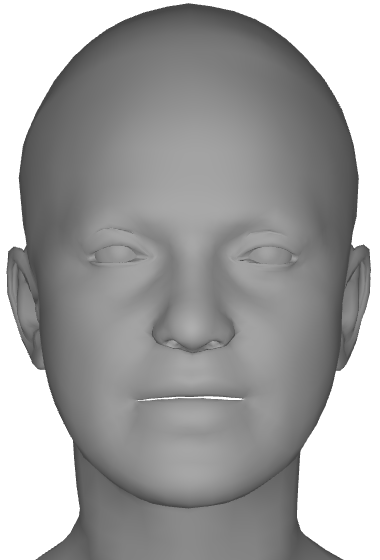}
    \end{subfigure}

        \centering
    \begin{subfigure}{0.075\textwidth}
        \centering
        \includegraphics[width=\linewidth]{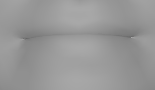}
    \end{subfigure}
    \hfill
    \begin{subfigure}{0.075\textwidth}
        \centering
        \includegraphics[width=\linewidth]{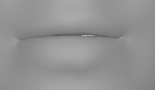}
    \end{subfigure}
    \hfill
    \begin{subfigure}{0.075\textwidth}
        \centering
        \includegraphics[width=\linewidth]{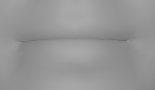}
    \end{subfigure}
    \hfill
    \begin{subfigure}{0.075\textwidth}
        \centering
        \includegraphics[width=\linewidth]{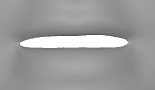}
    \end{subfigure}
    \hfill
    \begin{subfigure}{0.075\textwidth}
        \centering
        \includegraphics[width=\linewidth]{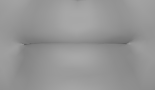}
    \end{subfigure}
    \hfill
    \begin{subfigure}{0.075\textwidth}
        \centering
        \includegraphics[width=\linewidth]{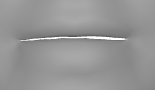}
    \end{subfigure}
    \vspace{-.2cm}
    \caption*{FaceFormer}

    \centering
    \begin{subfigure}{0.075\textwidth}
        \centering
        \includegraphics[width=\linewidth]{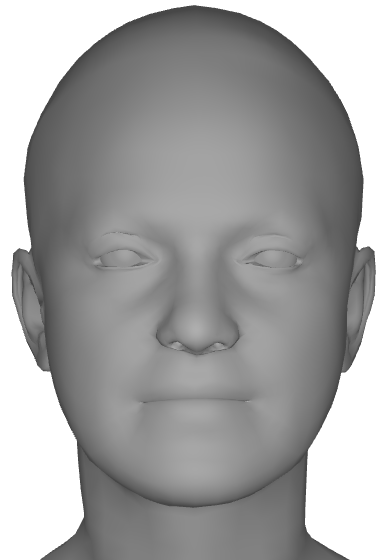}
    \end{subfigure}
    \hfill
    \begin{subfigure}{0.075\textwidth}
        \centering
        \includegraphics[width=\linewidth]{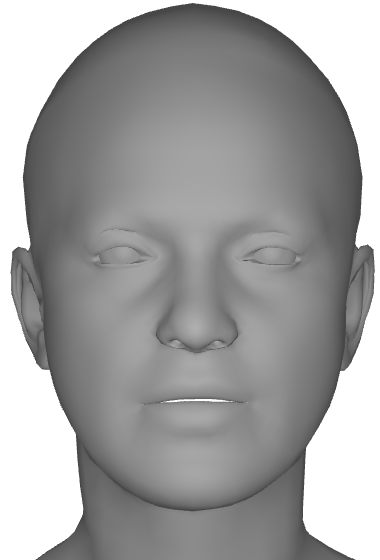}
    \end{subfigure}
    \hfill
    \begin{subfigure}{0.075\textwidth}
        \centering
        \includegraphics[width=\linewidth]{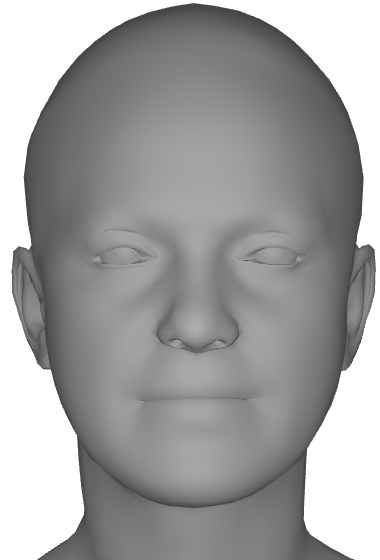}
    \end{subfigure}
    \hfill
    \begin{subfigure}{0.075\textwidth}
        \centering
        \includegraphics[width=\linewidth]{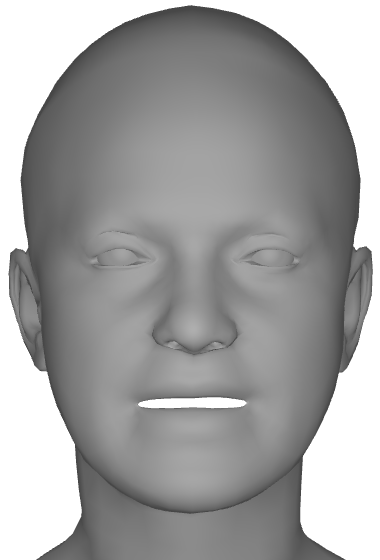}
    \end{subfigure}
    \hfill
    \begin{subfigure}{0.075\textwidth}
        \centering
        \includegraphics[width=\linewidth]{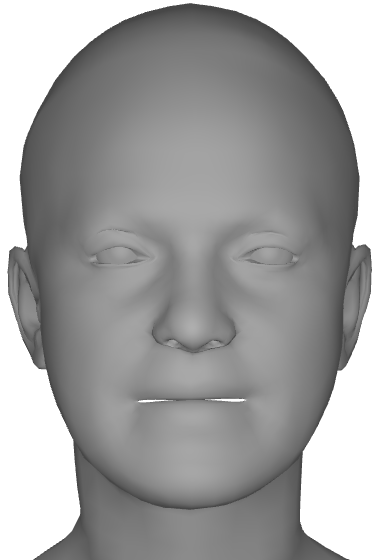}
    \end{subfigure}
    \hfill
    \begin{subfigure}{0.075\textwidth}
        \centering
        \includegraphics[width=\linewidth]{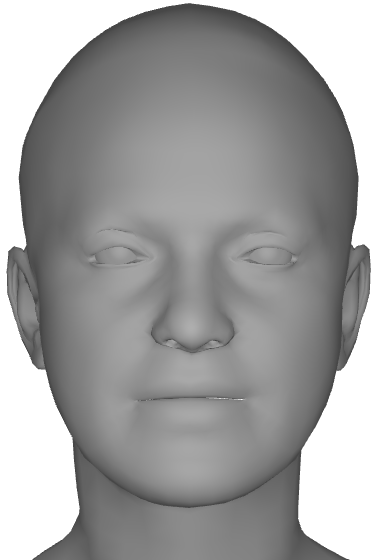}
    \end{subfigure}

    \centering
    \begin{subfigure}{0.075\textwidth}
        \centering
        \includegraphics[width=\linewidth]{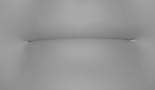}
    \end{subfigure}
    \hfill
    \begin{subfigure}{0.075\textwidth}
        \centering
        \includegraphics[width=\linewidth]{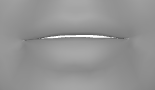}
    \end{subfigure}
    \hfill
    \begin{subfigure}{0.075\textwidth}
        \centering
        \includegraphics[width=\linewidth]{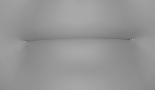}
    \end{subfigure}
    \hfill
    \begin{subfigure}{0.075\textwidth}
        \centering
        \includegraphics[width=\linewidth]{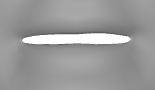}
    \end{subfigure}
    \hfill
    \begin{subfigure}{0.075\textwidth}
        \centering
        \includegraphics[width=\linewidth]{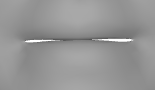}
    \end{subfigure}
    \hfill
    \begin{subfigure}{0.075\textwidth}
        \centering
        \includegraphics[width=\linewidth]{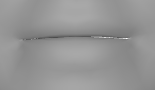}
    \end{subfigure}
    \vspace{-.2cm}
    \caption*{VOCA}

    \centering
    \begin{subfigure}{0.075\textwidth}
        \centering
        \includegraphics[width=\linewidth]{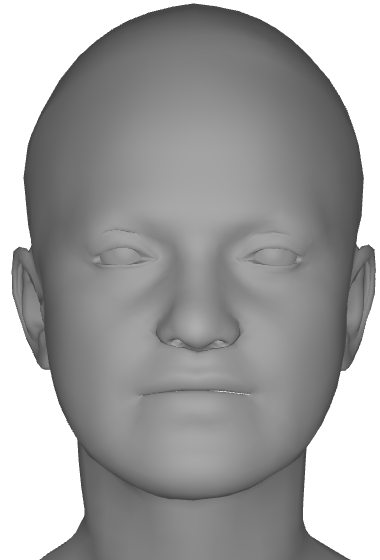}
    \end{subfigure}
    \hfill
    \begin{subfigure}{0.075\textwidth}
        \centering
        \includegraphics[width=\linewidth]{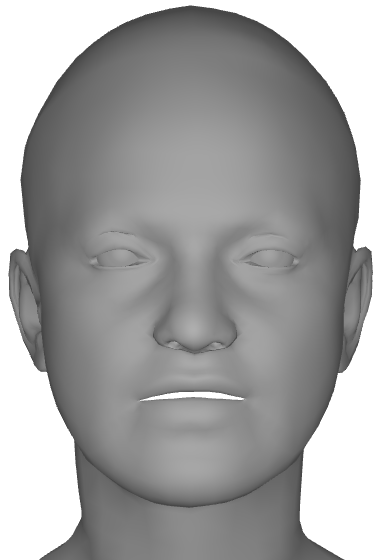}
    \end{subfigure}
    \hfill
    \begin{subfigure}{0.075\textwidth}
        \centering
        \includegraphics[width=\linewidth]{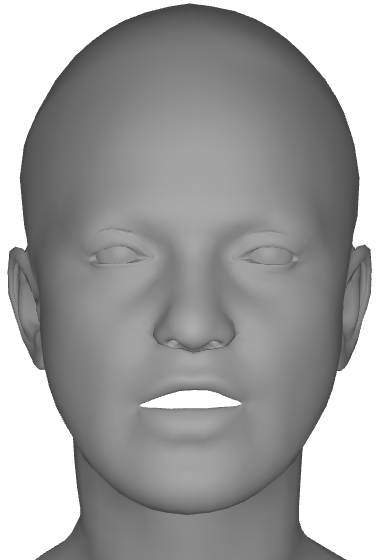}
    \end{subfigure}
    \hfill
    \begin{subfigure}{0.075\textwidth}
        \centering
        \includegraphics[width=\linewidth]{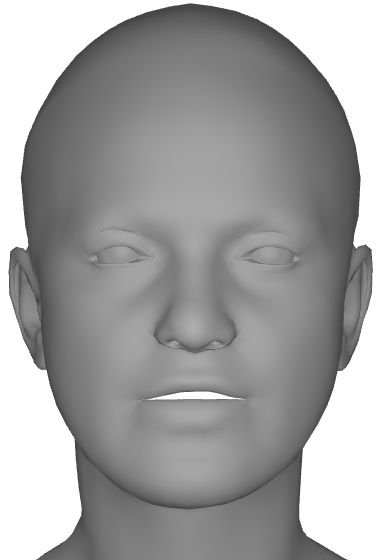}
    \end{subfigure}
    \hfill
    \begin{subfigure}{0.075\textwidth}
        \centering
        \includegraphics[width=\linewidth]{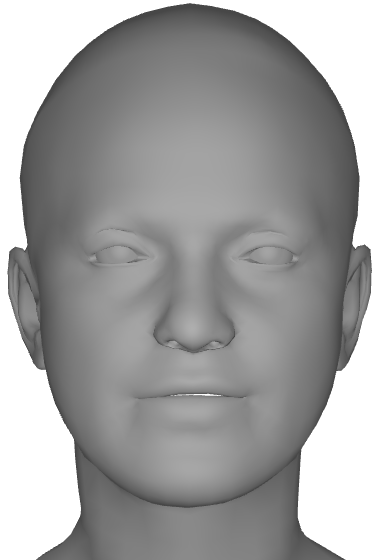}
    \end{subfigure}
    \hfill
    \begin{subfigure}{0.075\textwidth}
        \centering
        \includegraphics[width=\linewidth]{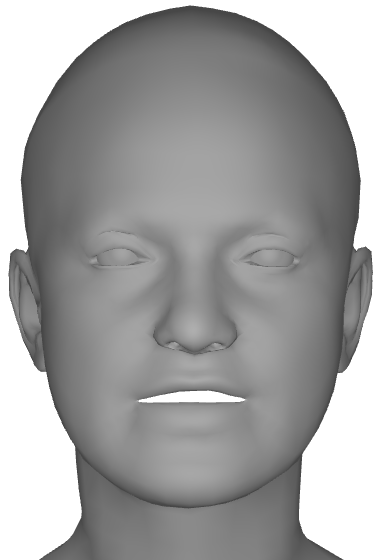}
    \end{subfigure}

    \centering
    \begin{subfigure}{0.075\textwidth}
        \centering
        \includegraphics[width=\linewidth]{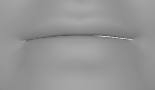}
    \end{subfigure}
    \hfill
    \begin{subfigure}{0.075\textwidth}
        \centering
        \includegraphics[width=\linewidth]{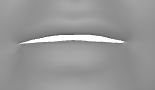}
    \end{subfigure}
    \hfill
    \begin{subfigure}{0.075\textwidth}
        \centering
        \includegraphics[width=\linewidth]{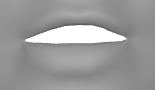}
    \end{subfigure}
    \hfill
    \begin{subfigure}{0.075\textwidth}
        \centering
        \includegraphics[width=\linewidth]{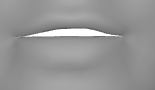}
    \end{subfigure}
    \hfill
    \begin{subfigure}{0.075\textwidth}
        \centering
        \includegraphics[width=\linewidth]{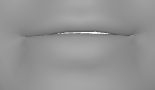}
    \end{subfigure}
    \hfill
    \begin{subfigure}{0.075\textwidth}
        \centering
        \includegraphics[width=\linewidth]{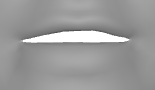}
    \end{subfigure}
    
    \caption{Face/lip motions at 6 time points for Sentence "Whoever cooperates in finding Nan's cameo will be rewarded." From the 2nd frame, OT-Talk shows the best synchronization with the Ground Truth, especially for the last 3 frames.}
    
\end{figure}

\begin{figure}[!htbp]
    
    \begin{subfigure}{.45\textwidth}
        \centering
        \includegraphics[width=1\linewidth]{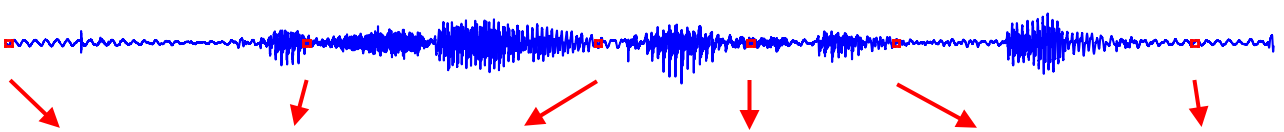}
    \end{subfigure}
    \vspace{-.3cm}
    \caption*{\textbf{Ground Truth}}
    \centering
    \begin{subfigure}{0.075\textwidth}
        \centering
        \includegraphics[width=\linewidth]{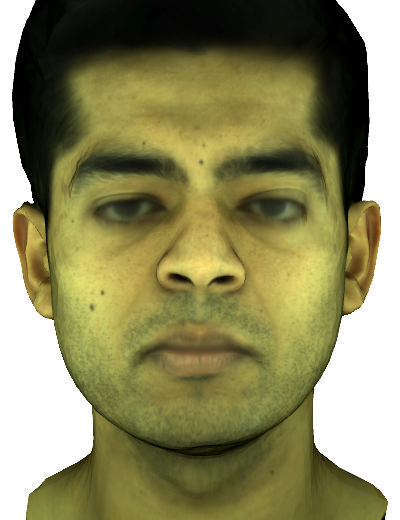}
    \end{subfigure}
    \hfill
    \begin{subfigure}{0.075\textwidth}
        \centering
        \includegraphics[width=\linewidth]{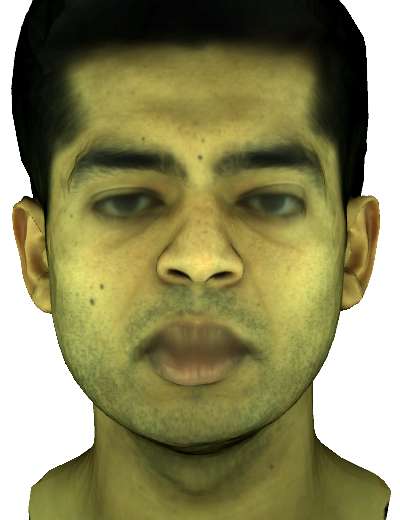}
    \end{subfigure}
    \hfill
    \begin{subfigure}{0.075\textwidth}
        \centering
        \includegraphics[width=\linewidth]{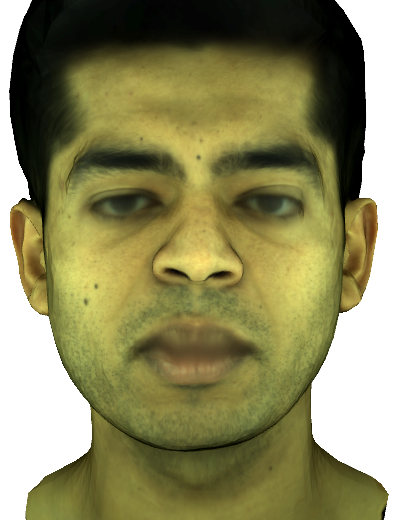}
    \end{subfigure}
    \hfill
    \begin{subfigure}{0.075\textwidth}
        \centering
        \includegraphics[width=\linewidth]{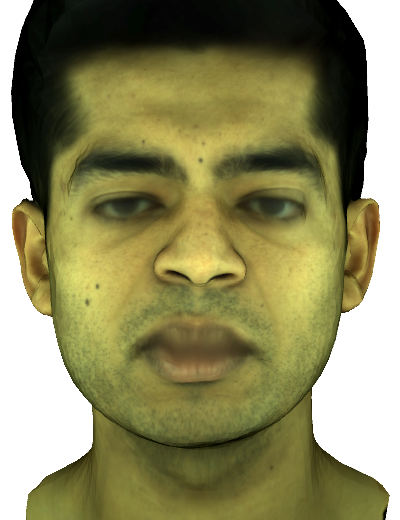}
    \end{subfigure}
    \hfill
    \begin{subfigure}{0.075\textwidth}
        \centering
        \includegraphics[width=\linewidth]{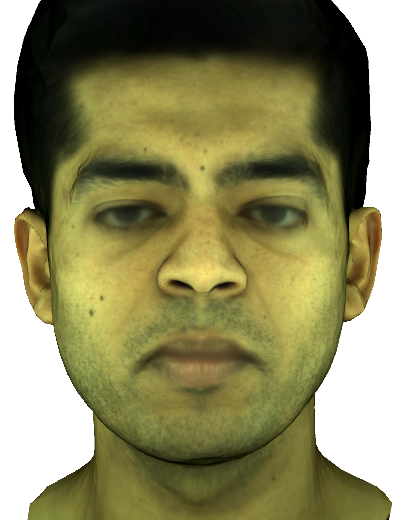}
    \end{subfigure}
    \hfill
    \begin{subfigure}{0.075\textwidth}
        \centering
        \includegraphics[width=\linewidth]{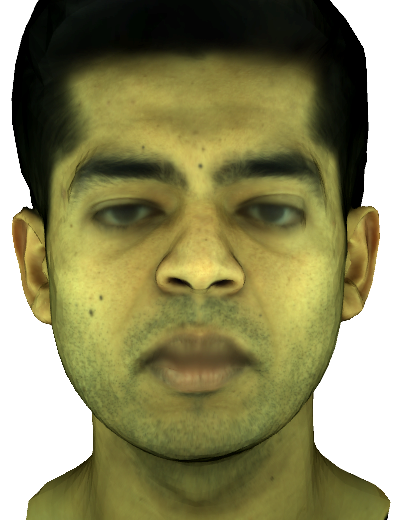}
    \end{subfigure}

    \centering
    \begin{subfigure}{0.075\textwidth}
        \centering
        \includegraphics[width=\linewidth]{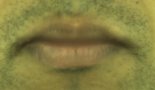}
    \end{subfigure}
    \hfill
    \begin{subfigure}{0.075\textwidth}
        \centering
        \includegraphics[width=\linewidth]{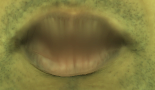}
    \end{subfigure}
    \hfill
    \begin{subfigure}{0.075\textwidth}
        \centering
        \includegraphics[width=\linewidth]{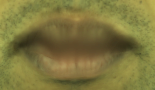}
    \end{subfigure}
    \hfill
    \begin{subfigure}{0.075\textwidth}
        \centering
        \includegraphics[width=\linewidth]{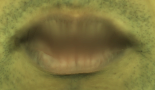}
    \end{subfigure}
    \hfill
    \begin{subfigure}{0.075\textwidth}
        \centering
        \includegraphics[width=\linewidth]{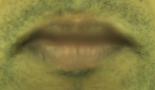}
    \end{subfigure}
    \hfill
    \begin{subfigure}{0.075\textwidth}
        \centering
        \includegraphics[width=\linewidth]{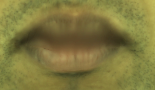}
    \end{subfigure}
    \caption*{OT-Talk}

    \centering
    \begin{subfigure}{0.075\textwidth}
        \centering
        \includegraphics[width=\linewidth]{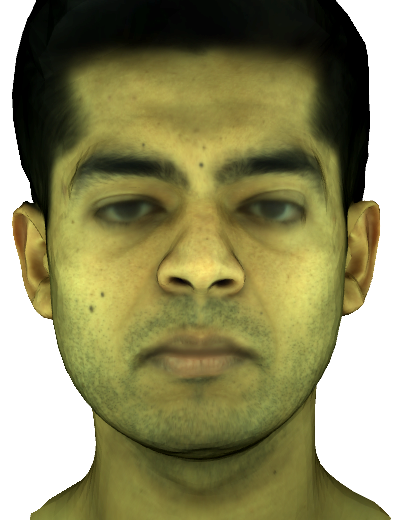}
    \end{subfigure}
    \hfill
    \begin{subfigure}{0.075\textwidth}
        \centering
        \includegraphics[width=\linewidth]{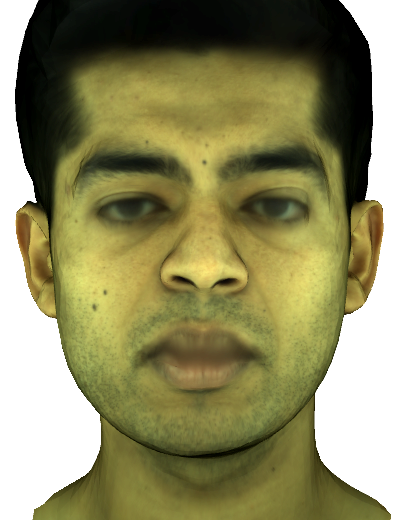}
    \end{subfigure}
    \hfill
    \begin{subfigure}{0.075\textwidth}
        \centering
        \includegraphics[width=\linewidth]{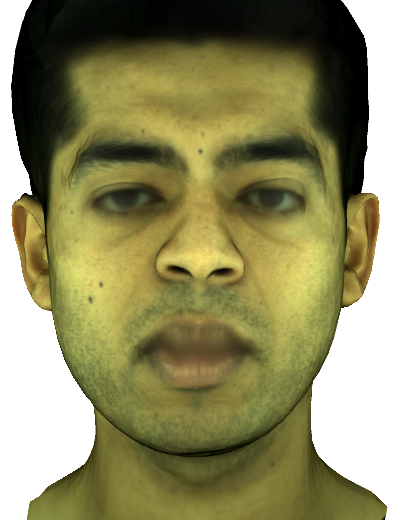}
    \end{subfigure}
    \hfill
    \begin{subfigure}{0.075\textwidth}
        \centering
        \includegraphics[width=\linewidth]{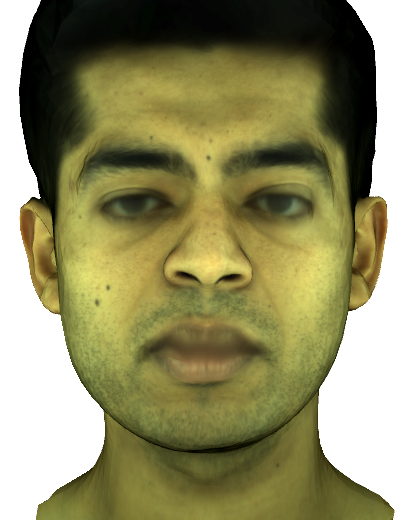}
    \end{subfigure}
    \hfill
    \begin{subfigure}{0.075\textwidth}
        \centering
        \includegraphics[width=\linewidth]{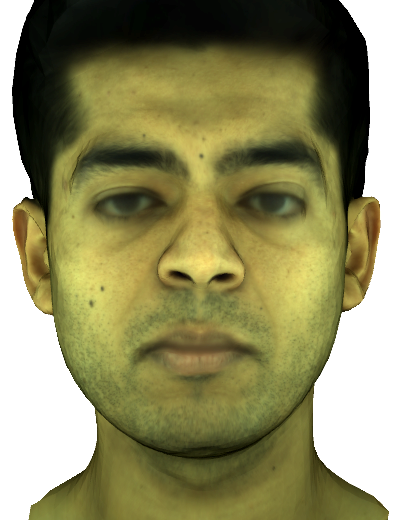}
    \end{subfigure}
    \hfill
    \begin{subfigure}{0.075\textwidth}
        \centering
        \includegraphics[width=\linewidth]{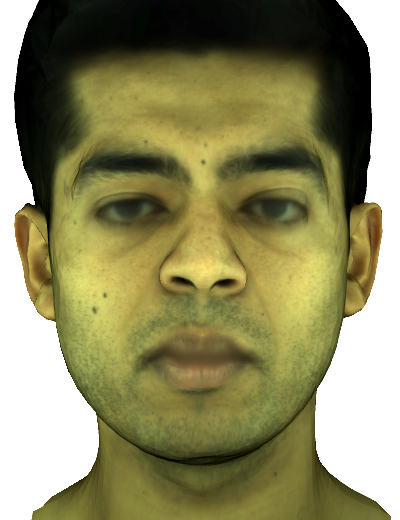}
    \end{subfigure}
    
    \centering
    \begin{subfigure}{0.075\textwidth}
        \centering
        \includegraphics[width=\linewidth]{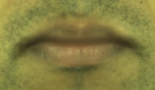}
    \end{subfigure}
    \hfill
    \begin{subfigure}{0.075\textwidth}
        \centering
        \includegraphics[width=\linewidth]{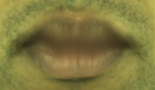}
    \end{subfigure}
    \hfill
    \begin{subfigure}{0.075\textwidth}
        \centering
        \includegraphics[width=\linewidth]{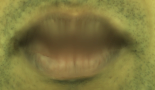}
    \end{subfigure}
    \hfill
    \begin{subfigure}{0.075\textwidth}
        \centering
        \includegraphics[width=\linewidth]{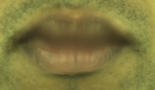}
    \end{subfigure}
    \hfill
    \begin{subfigure}{0.075\textwidth}
        \centering
        \includegraphics[width=\linewidth]{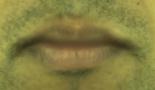}
    \end{subfigure}
    \hfill
    \begin{subfigure}{0.075\textwidth}
        \centering
        \includegraphics[width=\linewidth]{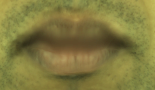}
    \end{subfigure}
    \caption*{CodeTalker}

    \centering
    \begin{subfigure}{0.075\textwidth}
        \centering
        \includegraphics[width=\linewidth]{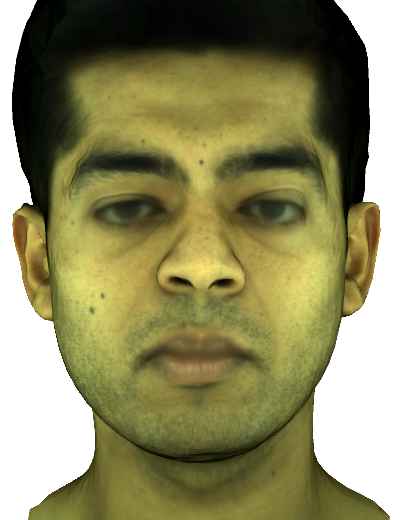}
    \end{subfigure}
    \hfill
    \begin{subfigure}{0.075\textwidth}
        \centering
        \includegraphics[width=\linewidth]{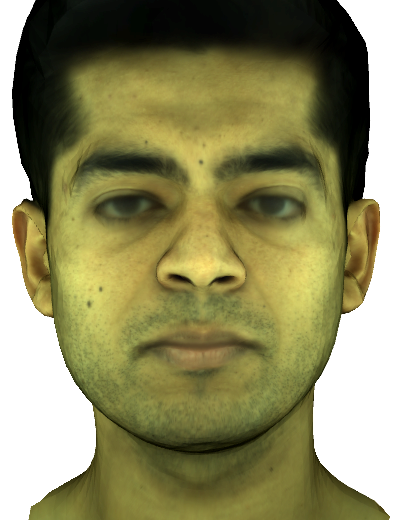}
    \end{subfigure}
    \hfill
    \begin{subfigure}{0.075\textwidth}
        \centering
        \includegraphics[width=\linewidth]{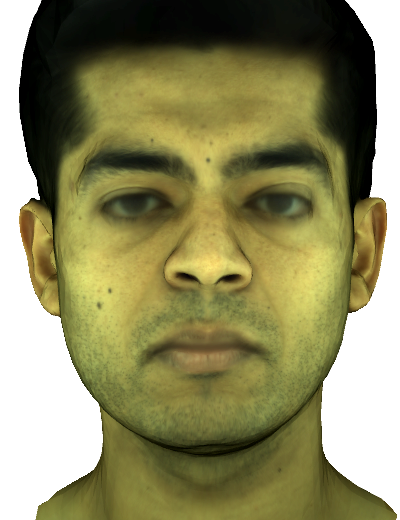}
    \end{subfigure}
    \hfill
    \begin{subfigure}{0.075\textwidth}
        \centering
        \includegraphics[width=\linewidth]{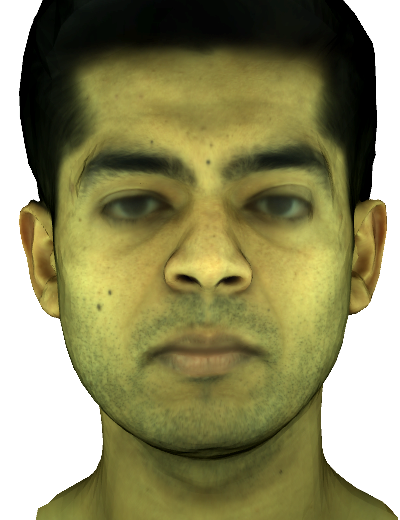}
    \end{subfigure}
    \hfill
    \begin{subfigure}{0.075\textwidth}
        \centering
        \includegraphics[width=\linewidth]{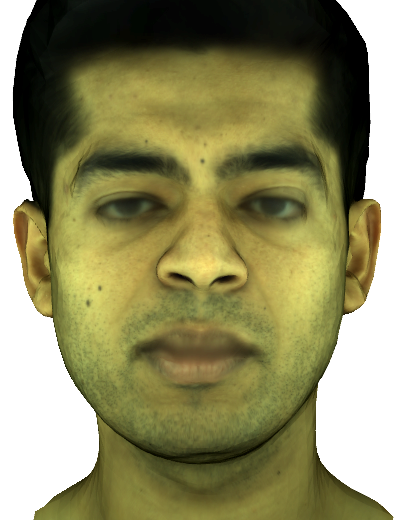}
    \end{subfigure}
    \hfill
    \begin{subfigure}{0.075\textwidth}
        \centering
        \includegraphics[width=\linewidth]{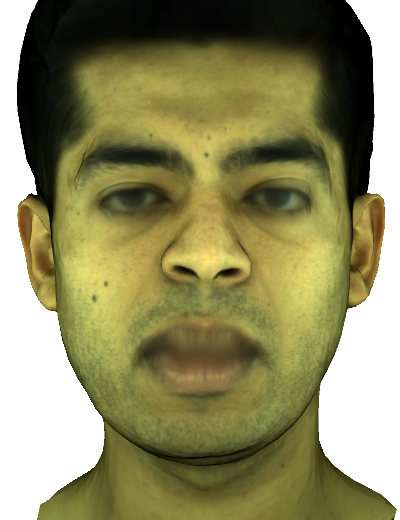}
    \end{subfigure}

    \centering
    \begin{subfigure}{0.075\textwidth}
        \centering
        \includegraphics[width=\linewidth]{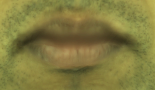}
    \end{subfigure}
    \hfill
    \begin{subfigure}{0.075\textwidth}
        \centering
        \includegraphics[width=\linewidth]{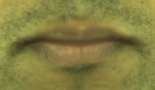}
    \end{subfigure}
    \hfill
    \begin{subfigure}{0.075\textwidth}
        \centering
        \includegraphics[width=\linewidth]{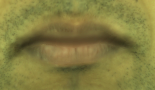}
    \end{subfigure}
    \hfill
    \begin{subfigure}{0.075\textwidth}
        \centering
        \includegraphics[width=\linewidth]{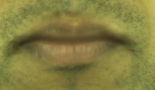}
    \end{subfigure}
    \hfill
    \begin{subfigure}{0.075\textwidth}
        \centering
        \includegraphics[width=\linewidth]{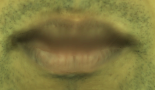}
    \end{subfigure}
    \hfill
    \begin{subfigure}{0.075\textwidth}
        \centering
        \includegraphics[width=\linewidth]{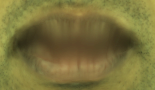}
    \end{subfigure}
    \caption*{FaceFormer}

    \centering
    \begin{subfigure}{0.075\textwidth}
        \centering
        \includegraphics[width=\linewidth]{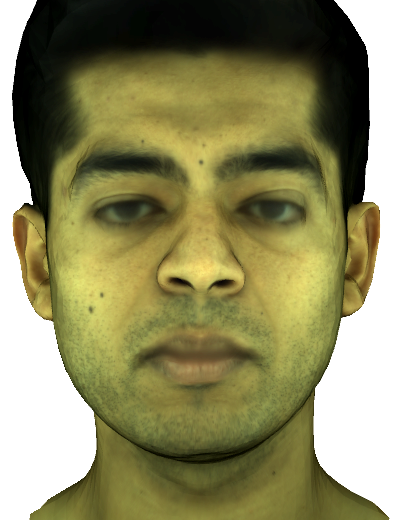}
    \end{subfigure}
    \hfill
    \begin{subfigure}{0.075\textwidth}
        \centering
        \includegraphics[width=\linewidth]{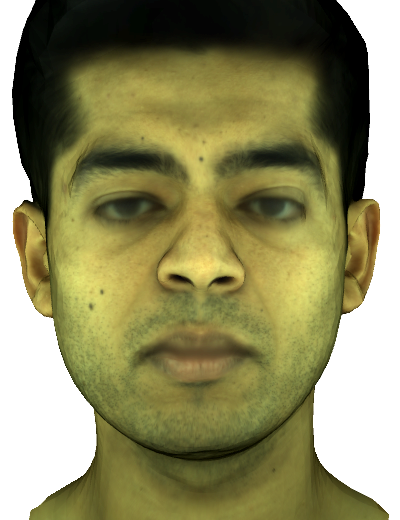}
    \end{subfigure}
    \hfill
    \begin{subfigure}{0.075\textwidth}
        \centering
        \includegraphics[width=\linewidth]{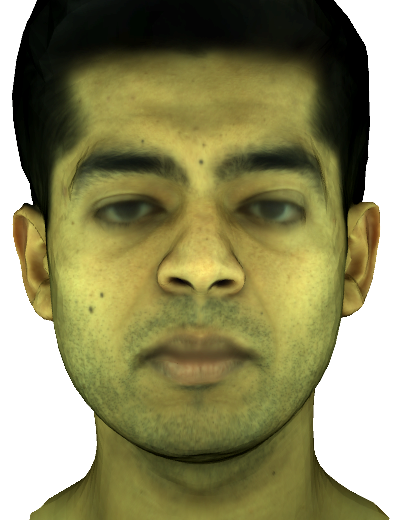}
    \end{subfigure}
    \hfill
    \begin{subfigure}{0.075\textwidth}
        \centering
        \includegraphics[width=\linewidth]{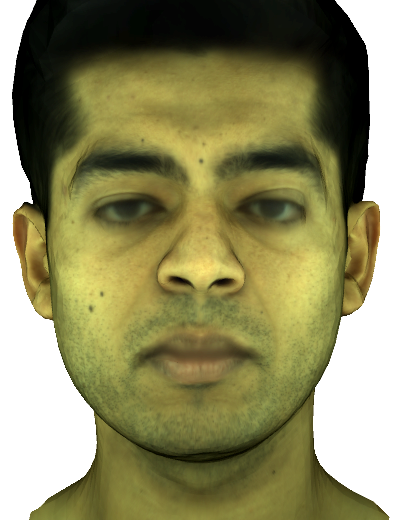}
    \end{subfigure}
    \hfill
    \begin{subfigure}{0.075\textwidth}
        \centering
        \includegraphics[width=\linewidth]{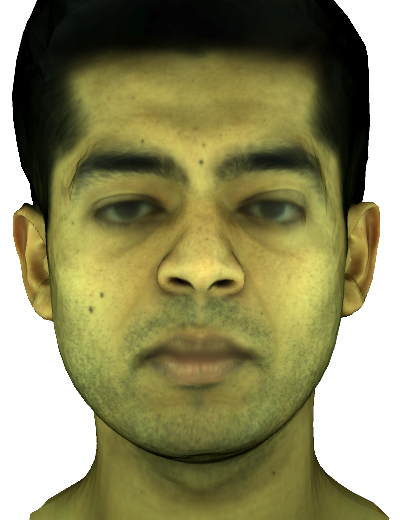}
    \end{subfigure}
    \hfill
    \begin{subfigure}{0.075\textwidth}
        \centering
        \includegraphics[width=\linewidth]{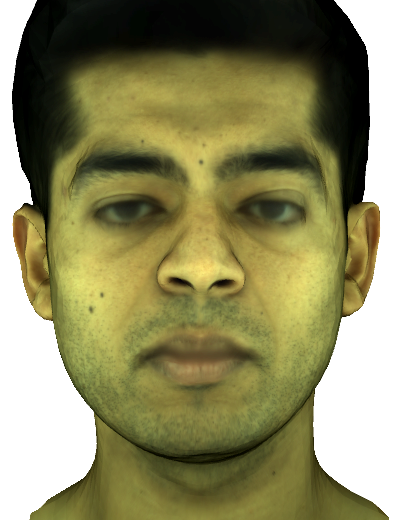}
    \end{subfigure}

    \centering
    \begin{subfigure}{0.075\textwidth}
        \centering
        \includegraphics[width=\linewidth]{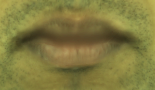}
    \end{subfigure}
    \hfill
    \begin{subfigure}{0.075\textwidth}
        \centering
        \includegraphics[width=\linewidth]{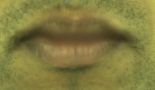}
    \end{subfigure}
    \hfill
    \begin{subfigure}{0.075\textwidth}
        \centering
        \includegraphics[width=\linewidth]{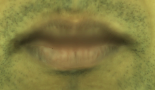}
    \end{subfigure}
    \hfill
    \begin{subfigure}{0.075\textwidth}
        \centering
        \includegraphics[width=\linewidth]{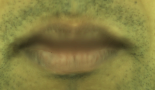}
    \end{subfigure}
    \hfill
    \begin{subfigure}{0.075\textwidth}
        \centering
        \includegraphics[width=\linewidth]{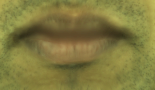}
    \end{subfigure}
    \hfill
    \begin{subfigure}{0.075\textwidth}
        \centering
        \includegraphics[width=\linewidth]{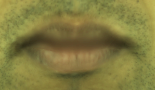}
    \end{subfigure}
    \caption*{VOCA}

    \centering
    \begin{subfigure}{0.075\textwidth}
        \centering
        \includegraphics[width=\linewidth]{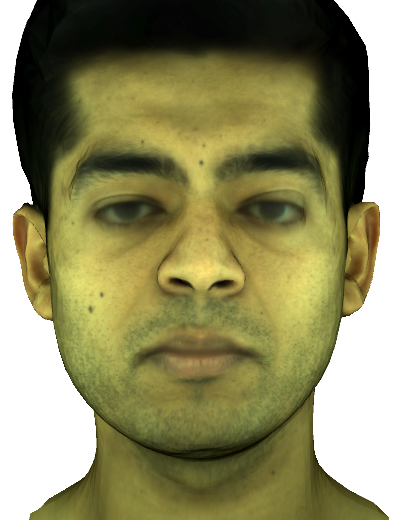}
    \end{subfigure}
    \hfill
    \begin{subfigure}{0.075\textwidth}
        \centering
        \includegraphics[width=\linewidth]{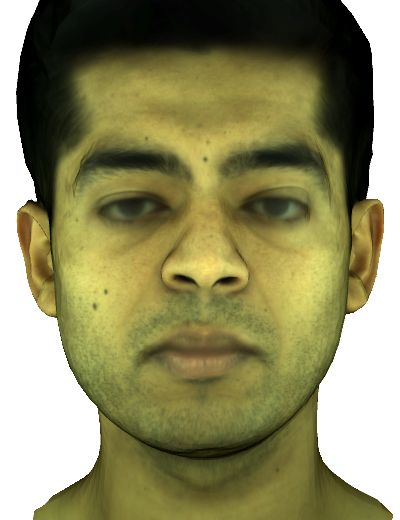}
    \end{subfigure}
    \hfill
    \begin{subfigure}{0.075\textwidth}
        \centering
        \includegraphics[width=\linewidth]{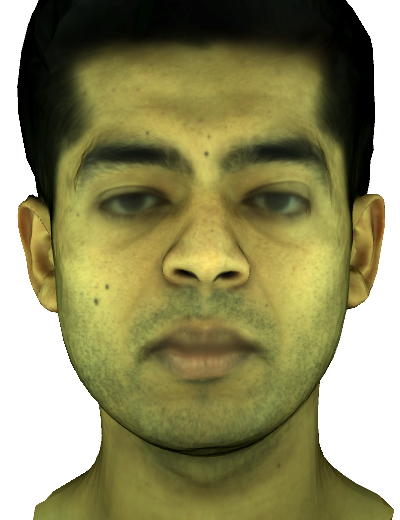}
    \end{subfigure}
    \hfill
    \begin{subfigure}{0.075\textwidth}
        \centering
        \includegraphics[width=\linewidth]{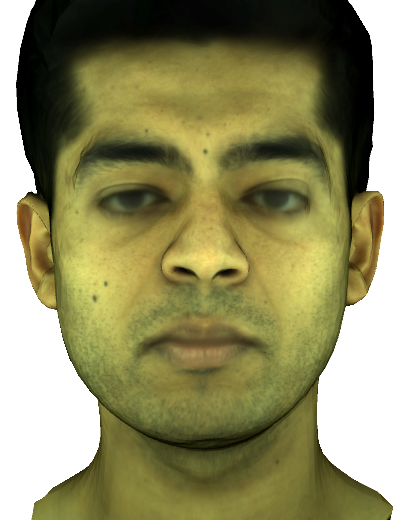}
    \end{subfigure}
    \hfill
    \begin{subfigure}{0.075\textwidth}
        \centering
        \includegraphics[width=\linewidth]{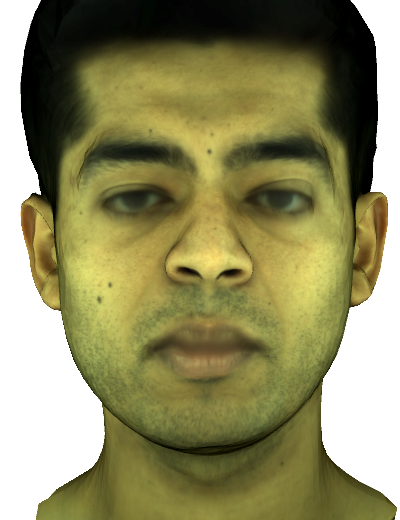}
    \end{subfigure}
    \hfill
    \begin{subfigure}{0.075\textwidth}
        \centering
        \includegraphics[width=\linewidth]{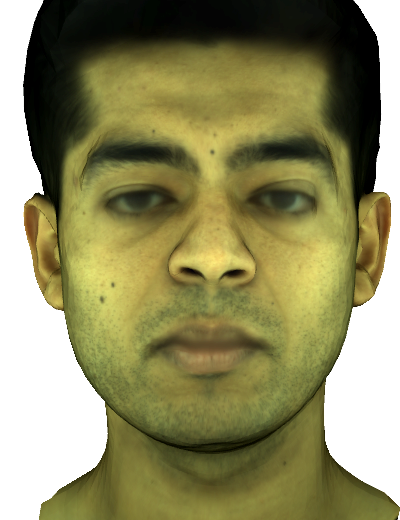}
    \end{subfigure}

    \centering
    \begin{subfigure}{0.075\textwidth}
        \centering
        \includegraphics[width=\linewidth]{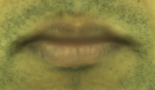}
    \end{subfigure}
    \hfill
    \begin{subfigure}{0.075\textwidth}
        \centering
        \includegraphics[width=\linewidth]{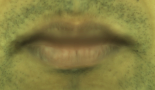}
    \end{subfigure}
    \hfill
    \begin{subfigure}{0.075\textwidth}
        \centering
        \includegraphics[width=\linewidth]{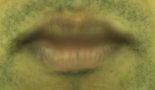}
    \end{subfigure}
    \hfill
    \begin{subfigure}{0.075\textwidth}
        \centering
        \includegraphics[width=\linewidth]{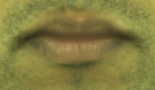}
    \end{subfigure}
    \hfill
    \begin{subfigure}{0.075\textwidth}
        \centering
        \includegraphics[width=\linewidth]{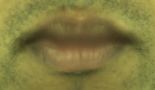}
    \end{subfigure}
    \hfill
    \begin{subfigure}{0.075\textwidth}
        \centering
        \includegraphics[width=\linewidth]{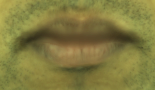}
    \end{subfigure}

    \caption{Face/lip motions at 6 time points for sentence "Did Shawn catch the big goose without help?" The 2nd, 3rd and 4th frames have obvious lip motions and only OT-Talk synchronizes with it at the corresponding time points.}
    
\end{figure}

\begin{figure}
    \centering
    \includegraphics[width=0.9\linewidth]{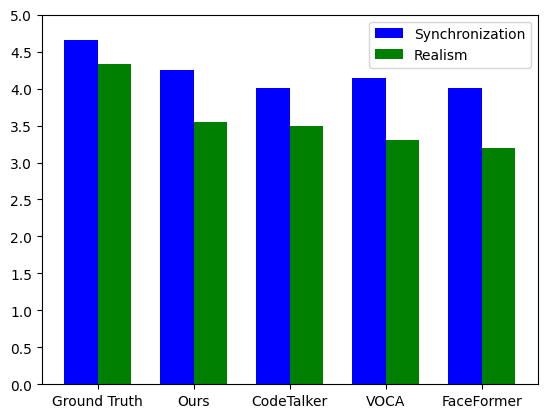}
    \vspace{-.3cm}
    \caption{Scores on VOCASET Dataset}
    \label{Scores on VOCASET Dataset}
\end{figure}

\begin{figure}
    \centering
    \includegraphics[width=0.9\linewidth]{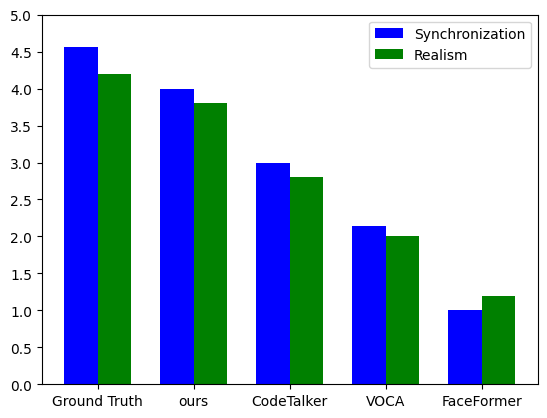}
    \vspace{-.3cm}
    \caption{Scores on Multiface Dataset}
    \label{Scores on Multiface Dataset}
\end{figure}

\subsection{Qualitative Results}
We present rendered meshes at certain frames pronouncing words with large lip/face motions to compare qualitative results. In Figure 3 we pick 4 words ('reward', 'hydride', 'plan', and 'what') spoken by 2 subjects from VOCASET. In Figure 4 we pick 4 words ('catch', 'class', 'how', and 'buy') spoken by 2 subjects from VOCASET. Since eye motions are not deterministic, all the approaches fail to synthesize the eye movement in the ground truth (last row in Figure 4). The 2 figures demonstrate that our results are the most close to the ground truths. In general, the qualitative performance is consistent with the quantitative results. 
Figures 5 and Figure 6 show the corresponding face/lip motions generated by different approaches during a sentence (Figure 5 picks a sentence from VOCASET and Figure 6 picks a sentence from Multiface). We uniformly pick six time points (red blocks in audio waves) and list the rendered faces. We find that our results are the most temporally synchronized with the ground truth because the lip motions take place at the closest time frames.

\subsection{User Perception Study}

We predict meshes for audio files and mesh templates in the test data, render them into images, and generate the corresponding videos. We use the same identity conditioning for VOCA, FaceFormer, and CodeTalker in the quantitative evaluation. We do this study separately on the 2 datasets. For VOCASET, each video clip consists of 9-10 sentences, lasting 40-50 seconds. For Multiface, each video clip consists of 4-5 sentences, lasting 15-20 seconds. The perceptual study involves 20 participants with good vision and hearing abilities. Participants are asked to provide scores (on a scale of 1 to 5) based on two aspects: (1) synchronization between lip movements and audio and (2) the realism of the synthesized results. In addition, participants are asked to perform an A over B test to determine if they prefer our approach or the baseline methods. Participants do not know by which method each video is generated. 

The scores are shown in Figures 7 and 8, while the A over B test results are presented in Tables 3 and 4. 
The results show a great difference between the datasets. Since VOCASET is a smoothed dataset, almost all approaches can achieve good performance, and the visual qualities of results generated by them are relatively close.
In sum, the user perception result is consistent with the quantitative comparison, but user perception feedback can be quite different on some video clips. E.g., some viewers do not like certain ground-truth video clips because the subject in the second row in Figure 3 always has extremely large lip motions.
Our approach shows a great advantage on Multiface dataset and baseline methods generate much worse results. FaceFormer tends to produce overly smooth results and failed to drive the movements of lips in many cases (resulting in the lowest score). The results of VOCA are not smooth while showing discontinuity between adjacent frames. CodeTalker has an advantage on synthesizing eye movement even compared with our approach but cannot keep consistent with lip dynamics in many cases.

\begin{table}[!htbp]
    \centering
    \begin{tabular}{c c c} 
         Method Pair&  Synchronization & Realism\\
         \hline
         ours / CodeTalker & 67.35 \% & 52.96\% \\
         ours / FaceFormer & 63.56\% & 59.43\% \\
         ours / VOCA & 52.71\% & 57.88\% \\
         ours / Ground Truth & 40.23\% & 38.07\% \\
         \hline
    \end{tabular}
    \caption{A Over B Perception Test on VOCASET Dataset}
\end{table}

\begin{table}[!htbp]
    \centering
    \begin{tabular}{c c c} 
         Method Pair&  Synchronization & Realism\\
         \hline
         ours / CodeTalker & 80.78\% & 80.56\% \\
         ours / FaceFormer & 99.06\% & 99.44\% \\
         ours / VOCA & 87.50\% & 87.78\% \\
         ours / Ground Truth & 47.11\% & 48.16\% \\
         \hline
    \end{tabular}
    \caption{A Over B Perception Test on Multiface Dataset}
\end{table}

\section{Conclusion and Limitation}

We introduce a novel approach OT-Talk for synthesizing 3D talking heads. It uses a ChebNet to enhance mesh representation and capture detailed geometric features, pre-trained Hubert and transformer encoder to extract audio features, and OT to evaluate predicted meshes. 
OT-Talk demonstrates significant improvements over baseline models in both quantitative and qualitative performance, providing better synchronization between facial movements and speech. The user perception study further shows the superiority of our approach, which is the most favored by the participants. 

However, our approach still has some limitations that need to be solved in future work. First, the control of the talking styles of different identities. Many previous methods use a one-hot label for talking style conditioning which is inaccurate and only able to generate a few number of talking styles. Our approach tries to embed the identity information with the latent code of a template mesh learned from a mesh encoder. Although the mesh encoder has a more powerful representation space, the datasets only capture a few subjects. Hence, we need more work on data collection. Second, the non-deterministic mapping from audio signals to face motions. Many facial expressions are not one-to-one determined by the speech input but are strongly related to it, which largely influence the perception quality of talking head animations. In addition to collecting audio-mesh data with richer and more diverse facial expressions or emotions, we will also explore diffusion models as FaceDiffuser \cite{stan2023facediffuser} or DiffPoseTalk \cite{sun2024diffposetalk} for this problem.



\bibliographystyle{ACM-Reference-Format}
\balance
\bibliography{sample-base}


\end{document}